\documentclass[fleqn,usenatbib]{mnras}

\usepackage{newtxtext,newtxmath}

\usepackage[T1]{fontenc}

\DeclareRobustCommand{\VAN}[3]{#2}
\let\VANthebibliography\thebibliography
\def\thebibliography{\DeclareRobustCommand{\VAN}[3]{##3}\VANthebibliography}

\usepackage{graphicx}	
\usepackage{amsmath}	
\usepackage{lscape}
\usepackage{setspace}
\usepackage{comment}
\usepackage{color, colortbl}
\usepackage{orcidlink}

\usepackage{xcolor}
\usepackage{chngcntr}

\newcommand{\GenevaObservatory}{Observatoire de Gen\`eve, Universit\'e de Gen\`eve, 51 chemin des Maillettes, 1290 Versoix, Switzerland}
\newcommand{\kms}{\ensuremath{\rm km\,s^{-1}}}
\newcommand{\cms}{\ensuremath{\rm cm\,s^{-1}}}
\newcommand{\ms}{\ensuremath{\rm m\,s^{-1}}}
\providecommand{\bjdtdb}{\ensuremath{\rm {BJD_{TDB}}}}

\providecommand{\msun}{\ensuremath{\,M_\Sun}}
\providecommand{\rsun}{\ensuremath{\,R_\Sun}}
\providecommand{\lsun}{\ensuremath{\,L_\Sun}}
\providecommand{\mj}{\ensuremath{\,M_{\rm J}}}

\providecommand{\rj}{\ensuremath{\,R_{\rm J}}}
\providecommand{\me}{\ensuremath{\,M_{\oplus}}}
\providecommand{\re}{\ensuremath{\,R_{\oplus}}}
\providecommand{\fave}{\langle F \rangle}
\providecommand{\fluxcgs}{10$^9$ erg s$^{-1}$ cm$^{-2}$}
\newcommand{\TESS}{\emph{TESS}}
\newcommand{\Ktwo}{\emph{K2}}
\newcommand{\calm}{\emph{CALM}}

\definecolor{new_color}{HTML}{000000}

\definecolor{html_black}{HTML}{000000}

\newcommand\bedit[1]{\textcolor{html_black}{#1}} 
\newcommand\beditr[1]{\textcolor{html_black}{#1}} 
\newcommand\beditrr[1]{\textcolor{html_black}{#1}} 

\newcommand\redit[1]{\textcolor{html_black}{#1}} 
\newcommand\rredit[1]{\textcolor{new_color}
{#1}}

\providecommand{\thisstar}{K2-167}
\providecommand{\thisplanet}{K2-167 b}

\providecommand{\olddrsrmsRAW}{\ensuremath{4.02}}

\providecommand{\olddrssigmaUS}{\ensuremath{2.04}}

\providecommand{\newdrsrmsRAW}{\ensuremath{3.01}}

\providecommand{\newdrssigmaAC}{\ensuremath{2.78}}
\providecommand{\newdrssigmaUS}{\ensuremath{2.41}}

\providecommand{\litperiod}{\ensuremath{9.978543}}

\providecommand{\bonomomass}{\ensuremath{6.5_{-1.5}^{+1.6}}}
\providecommand{\erikamass}{\ensuremath{7.0^{+1.7}_{-1.7}}} 
\providecommand{\calmmass}{\ensuremath{6.3_{-1.4}^{+1.4}}}

\providecommand{\erikaradius}{\ensuremath{2.33^{+0.17}_{-0.15}}}

\providecommand{\erikaK}{\ensuremath{1.88^{+0.49}_{-0.48}}}
\providecommand{\calmK}{\ensuremath{1.94^{+0.44}_{-0.44}}} 

\title[\redit{Characterization} of \thisplanet]{Characterization of \thisplanet\ and \rredit{CALM,} a new stellar activity mitigation method}

\author[Z. L. de Beurs et al.]{
Zo{\"e}\ L. de Beurs \orcidlink{0000-0002-7564-6047},$^{1,2}$\thanks{Corresponding author email: zdebeurs@mit.edu}
Andrew Vanderburg\orcidlink{0000-0001-7246-5438},$^{3}$
Erica Thygesen\orcidlink{0000-0002-9165-6245},$^{4, \dagger}$ 
Joseph E. Rodriguez\orcidlink{0000-0001-8812-0565},$^{4}$
Xavier Dumusque\orcidlink{0000-0002-9332-2011}, $^{5}$
\newauthor{Annelies Mortier\orcidlink{0000-0001-7254-4363},$^{6}$ Luca Malavolta\orcidlink{0000-0002-6492-2085}, $^{7, 8}$
Lars A. Buchhave\orcidlink{0000-0003-1605-5666},$^{9}$
Christopher J. Shallue\orcidlink{0000-0002-7585-9974}, $^{10}$} 
Sebastian Zieba\orcidlink{0000-0003-0562-6750},  $^{11, 12}$
\newauthor{Laura Kreidberg \orcidlink{0000-0003-0514-1147},$^{11}$}
John H. Livingston\orcidlink{0000-0002-4881-3620}, $^{13, 14, 15}$
\textcolor{new_color}{R. D. Haywood\orcidlink{0000-0001-9140-3574},$^{16, 17}$}
\textcolor{new_color}{David W. Latham \orcidlink{0000-0001-9911-7388},$^{10}$}
\newauthor{\textcolor{new_color}{Mercedes L\'opez-Morales}\orcidlink{0000-0003-3204-8183}, $^{11}$}
\textcolor{new_color}{Andr\'e M. Silva \orcidlink{0000-0003-4920-738X},$^{18, 19}$}
\\
$^{1}$Department of Earth, Atmospheric and Planetary Sciences, Massachusetts Institute of Technology, Cambridge, MA 02139, USA\\
$^{2}$NSF Graduate Research Fellow \& MIT Presidential Fellow \\
$^{3}$ Department of Physics and Kavli Institute for Astrophysics and Space Research, Massachusetts Institute of Technology, Cambridge, MA 02139, USA \\
$^{4}$ Center for Data Intensive and Time Domain Astronomy, Department of Physics and Astronomy, Michigan State University, East Lansing, MI 48824, USA \\
$^{5}$ \GenevaObservatory \\
$^{6}$ \bedit{School of Physics \& Astronomy, University of Birmingham, Edgbaston, Birmingham, B15 2TT, UK} \\
$^{7}$ Dipartimento di Fisica e Astronomia ``Galileo Galilei'', Universitá di Padova, Vicolo del l’Osservatorio 3, I-35122 Padova, Italy \\
$^{8}$ INAF - Osservatorio Astronomico di Padova, Vicolo dell’Osservatorio 5, Padova, 35122, Italy \\
$^{9}$ DTU Space, National Space Institute, Technical University of Denmark, Elektrovej 328, DK-2800 Kgs. Lyngby, Denmark \\
$^{10}$ Center for Astrophysics $\mid$ Harvard \& Smithsonian, 60 Garden St, Cambridge, MA 02138, USA \\
$^{11}$ Max-Planck-Institut f{\"u}r Astronomie, K{\"o}nigstuhl 17, D-69117
Heidelberg, Germany. \\
$^{12}$ Leiden Observatory, Leiden University, Niels Bohrweg 2, 2333CA Leiden, The Netherlands \\
$^{13}$ Astrobiology Center, 2-21-1 Osawa, Mitaka, Tokyo 181-8588, Japan \\
$^{14}$ National Astronomical Observatory of Japan, 2-21-1 Osawa, Mitaka, Tokyo 181-8588, Japan \\
$^{15}$ Astronomical Science Program, Graduate University for Advanced Studies, SOKENDAI, 2-21-1, Osawa, Mitaka, Tokyo, 181-8588, Japan \\
\textcolor{new_color}{$^{16}$ Astrophysics Group, University of Exeter, Exeter EX4 2QL, UK} \\
\textcolor{new_color}{$^{17}$ STFC Ernest Rutherford Fellow} \\
\textcolor{new_color}{$^{18}$ Departamento de Fisica e Astronomia, Faculdade de Ciencias, Universidade do Porto, Rua do Campo Alegre, 4169-007 Porto, Portugal} \\
\textcolor{new_color}{$^{19}$ Instituto de Astrofisica e Ciencias do Espaco, Universidade do Porto, CAUP, Rua das Estrelas, 4150-762 Porto, Portugal} \\
$^{\dagger}$ Quad Fellow}

\date{Accepted XXX. Received YYY; in original form ZZZ}

\pubyear{2015}

\begin{document}
\label{firstpage}
\pagerange{\pageref{firstpage}--\pageref{lastpage}}
\maketitle

\begin{abstract}
We report precise radial velocity (RV) observations of HD 212657 (= \thisstar), a star shown by \Ktwo\ to host a transiting sub-Neptune-sized planet in a 10 day orbit. Using Transiting Exoplanet Survey Satellite (\TESS) \redit{photometry}, we refined \redit{the} planet parameters, especially the orbital period. We collected 74 precise RVs with the HARPS-N spectrograph between August 2015 and October 2016. Although this planet was first found \redit{to transit} in 2015 and validated in 2018, excess RV scatter originally limited \redit{mass measurements. Here,} we \redit{measure} a mass \redit{by} taking advantage of reductions in scatter from updates to the HARPS-N Data Reduction System (2.3.5) and \redit{our} new activity mitigation method called CCF Activity Linear Model (\calm)\redit{, which} uses activity-induced line shape changes in the spectra without requiring timing information. Using the \calm\ framework, we performed a joint fit with RVs and transits using EXOFASTv2 and find $M_p = $ \calmmass \me\ and $R_p = $ \erikaradius \re, which places \thisplanet\ at the upper edge of the radius valley. We also find hints of a secondary companion at a $\sim$ 22 day period, but \redit{confirmation requires additional RVs. Although characterizing lower-mass planets like \thisplanet\ is often impeded by stellar variability, these systems} especially help probe the \redit{formation} physics (i.e. photoevaporation, core-powered mass loss) of the radius valley. In the future, \redit{\calm} or similar techniques could be widely applied to FGK-type stars, \redit{help} characterize a population of exoplanets \redit{surrounding} the radius valley, and \redit{further our understanding} of their formation.
\end{abstract}

\begin{keywords}
exoplanets -- techniques: radial velocities -- planets and satellites : terrestrial planets -- stars : activity -- methods : statistical
\end{keywords}



\section{Introduction}
\beditr{The study of exoplanets has the ultimate goal of understanding the formation and evolution of planets. In our own solar system, we have a class of dense smaller rocky planets with iron cores and a class of large gas giants that have \redit{extended} H/He envelopes. However, the HARPS spectrograph \citep{mayor2003} and \textit{Kepler} mission \citep{Borucki2010} found hundreds of planets \citep{mayor2011, batalha2013} that fall between these two categories and with it brought many new questions: What is the composition of these in-between planets? Does liquid water exist on their surface? Do they have gaseous envelopes? If so, do their envelopes mainly consist of H/He? How did this class of planets form and evolve? And why do we not see any examples of them in our own solar system? To answer these questions, we need to characterize these types of systems and constrain their possible compositions through precise radius and mass measurements.}

\beditr{Precise radial velocity (RV) observations revealed that \redit{transiting} planets larger than 1.6 \re\ are typically inconsistent with rocky compositions \citep{rogers2015}, so these intermediate planets are often sub-divided into two classes: super-Earths ($1 \re $  \textcolor{new_color}{$\lesssim$} $R_p  \lesssim 1.6 \re $) and sub-Neptunes ($1.6 \re \lesssim R_p  \lesssim 4 \re $). A gap in the occurrence of planets with radii between these two populations was predicted by several researchers based on the effect of photoevaporation \citep{2013Lopez, 2013Owen, 2014Jin}. This gap is often described as the ``radius valley'' or ``radius gap'' and describes a \redit{paucity} of planets between 1.5 to 2 \re. This bimodality in the occurence \redit{rates} of small planets has been detected \citep{2017Fulton} and further refined and characterized using parallax and asteroseismic measururements \citep{2018Berger, 2018VanEylen, 2020Petigura}. }

\beditr{Recently, \citet{2022Petigura} investigated the location of the radius gap as a function of stellar mass, metallicity, age, and orbital period and found that the radius gap location follows $R_p \propto P^m$ where $m = -0.10 \pm 0.03$.  \citet{dattilo2023} also found that the location of the radius gap varies depending on stellar type for FGK stars. However, the exact underlying physics that explains the origins of the radius gap is still somewhat debated \citep[e.g.][]{ginzburg, gupta, lee} and determining which mechanism is primarily responsible requires precise characterization of a larger population of exoplanets \citep{rogers}. Precise radius and mass measurements of planets at the boundaries of the radius valley can especially help refine our understanding of the physics by constraining the bulk planetary composition and inferring either the presence or absence of a thick volatile envelope.}

\beditr{Unfortunately, the difficulty of \redit{modeling stellar variability in RV} observations has limited the study of these important systems. In particular, spurious RV shifts due to stellar activity signals can mask or mimic planetary signals. In the past decade, stellar activity has been identified as the limiting factor in RV precision across the HR diagram \citep{national2018exoplanet, 2020arXiv200513386H}, but it is especially problematic for young stellar hosts, including those where planets might still be undergoing photoevaporation. Finding new ways to mitigate stellar activity is critical for our understanding of planetary composition and evolution.} 

Stellar activity mitigation methods have traditionally focused on using indicators of the level of stellar activity, which can either be focused on measuring the level of magnetic activity on a star \citep{Boisse2009, Dumusque2011, 2007A&A...474..293B, 2014Sci...345..440R} or focused on detecting the presence (or lack thereof) of spots and plagues on the surface \citep{Figueira2013, Milbourne2021, Wise2022}. \beditr{The activity indicators that correlate with magnetic activity include spectral features such as the emission in the core of Ca II H\&K lines \citep{Saar1998, Meunier2013, Saar2000}, the H$_\alpha$ line \citep{2007A&A...474..293B, 2014Sci...345..440R}, and the Ca infrared triplet.} \redit{Recently, photospheric unsigned magnetic flux has been shown to be strongly correlated with RV perturbations caused by solar surface activity \citep{Haywood2022} and new methods for deriving proxies for magnetic flux using least squares deconvolution similarly show some of the strongest correlations among activity indicators \citep{Lienhard2023}.} Beyond these types of activity indicators, properties in cross-correlation function (CCFs) have been used to decorrelate stellar activity from RVs. CCFs are computed by cross-correlating the spectra with a binary mask which contains delta functions at the location of spectral lines. Conceptually, a CCF represents an average of all line shapes in a star's spectrum and is therefore sensitive to line shape changes that persist in most stellar lines. The line shape changes introduced by stellar activity that cause asymmetry in the CCF are commonly measured using a CCF Bisector Inverse Span \citep[e.g.][]{2001A&A...379..279Q, Lovis2011} or the FWHM of the CCF \citep[e.g.][]{Queloz2009, Hebrard2010}. Although the bisector and FWHM have the advantage that they can easily be computed and provide only two free parameters in RV models, much of the information present in the CCF is lost by only using these metrics to measure line shape changes. In addition, these traditional stellar activity indicators have been insufficient to reach the sub \ms\ RV measurement precision necessary to detect earth-twins.

There have been recent developments in using advanced flexible models like Gaussian Process (GP) regression to model quasi-periodic stellar activity signals and reveal planet signals \citep{2014MNRAS.443.2517H, 2015MNRAS.452.2269R, 2017arXiv171101318J, 2018A&A...614A.133D, Barragan2019,Gilbertson2020, Nicholson2022, Barragan2023,Tran2023}. However, even with the current \bedit{state-of-the-art} stellar variability mitigation techniques \citep{Zhao2022}, our RV measurement precision must still improve by an order of magnitude to detect the 10 \cms\ signals induced by Earth-mass exoplanets in the habitable zones of Sun-like stars. To work towards achieving this goal, we can first focus on precise mass measurements of slightly more massive planets like super-Earths and sub-Neptunes. This requires us to \bedit{continue} develop\bedit{ing new}  methods to mitigate the impact of stellar activity on exoplanet \bedit{mass measurements}. 

Inspired by the successful implementation of a neural network stellar activity mitigation technique on solar observations \citep{deBeurs2022b} and other uses of machine and deep learning methods for modeling stellar activity \citep{Perger2023, Colwell2023}, we designed a simplified method \bedit{called CCF Activity Linear Model (\calm)}, which can be applied to \redit{other} stars \bedit{with smaller RV datasets} observed at nighttime. Our method focuses on exploiting the information content present in the CCFs and traces the line shape changes introduced by stellar activity to separate \beditr{D}oppler reflex motion from signals from the star. \beditr{We have already successfully applied this method to RVs for four stars observed with the EXPRES spectrograph and reduced the scatter by \redit{about a factor of two} for the more active stars \citep{Zhao2022}. In this paper, we apply this same method to} one star observed by HARPS-N: \thisstar.

\thisplanet\ is a planet located just above the radius valley and \redit{this} transiting sub-Neptune \redit{was first found} by the \Ktwo\ mission \citep{2016ApJS..222...14V, 2018AJ....155..136M}. \redit{T}he planetary signal was re-detected \citep{2020AJ....160..209I} by \redit{the Transiting Exoplanet Survey Satellite }\citep[\TESS;][]{2015JATIS...1a4003R}, and later again in its first extened mission \citep{Thygesen2023}.  HARPS-N observations of \thisstar, \redit{a relatively active, yet slowly rotating F7-type star,} began in late 2015, but in the intervening years, stellar activity had prevented a confident detection of the planet's mass. \beditrr{Recently, \citet{Bonomo2023} analyzed the \thisstar\ system as part of a larger sample and reported a mass of \bonomomass\ \me. Here, we perform a more in-depth analysis of the RV \redit{time series} than this previous work (which was included as part of a \redit{catalog} paper).} \redit{Using our \calm\ framework, we performed a joint fit of the transits and RVs, constraining the radius and mass, and enabling future studies of its composition which may provide insight into its formation.}

Our paper is organized as follows. In Section \ref{data}, we describe the the spectroscopic observations from HARPS-N and the photometric observations from \Ktwo\ and \TESS. \beditr{In Section \ref{rv_analysis}, we describe our development of the \calm\ stellar activity mitigation method,  how we prevent overfitting on our data, and introduce a new metric to measure RV scatter.} In Section \beditr{\ref{system_parameters}}, we describe the transit, spectroscopic, and radial velocity analysis and result\beditr{ing system parameters of \thisstar}. In Section \redit{\ref{discussionconclusions}}, we discuss the implications of these results and we conclude in Section \ref{Conclusion}.

\section{Data}\label{data}

\subsection{HARPS-N Spectroscopy}
\thisstar\ was observed with the HARPS-N Spectrograph on the 3.58m Telescopio Nazionale Galileo (TNG).  Designed specifically to \redit{yield precise} radial velocity measurements, HARPS-N is a vacuum-enclosed cross-dispersed echelle spectrograph with  temperature and pressure stabilization \citep{2012SPIE.8446E..1VC}. We collected 76 precise radial velocity observations of \thisstar\ (Table \ref{harpsrvs}) with typical integration times of 15 minutes\footnote{All observations had 15 minute integration times except for one 20 minute exposure on 26 September 2016 and two 30-minute observations on 28 November 2015 and 9 July 2016 respectively.} with the HARPS-N spectrograph between August 2015 and October 2016. For the first season of observations (20 August 2015 to 27 December 2015), a ThAr lamp served as a calibration source. For the second season (20 June 2016 to 12 October 2016), a stabilized Fabry Perot \redit{was} used as the calibration source. Most of the time, we obtained only one observation per night, but between July and November 2016, we often obtained two spectra per night, separated by a few hours, in order to average radial velocity variations due to granulation.

\begin{figure}
    \includegraphics[width=1\linewidth]{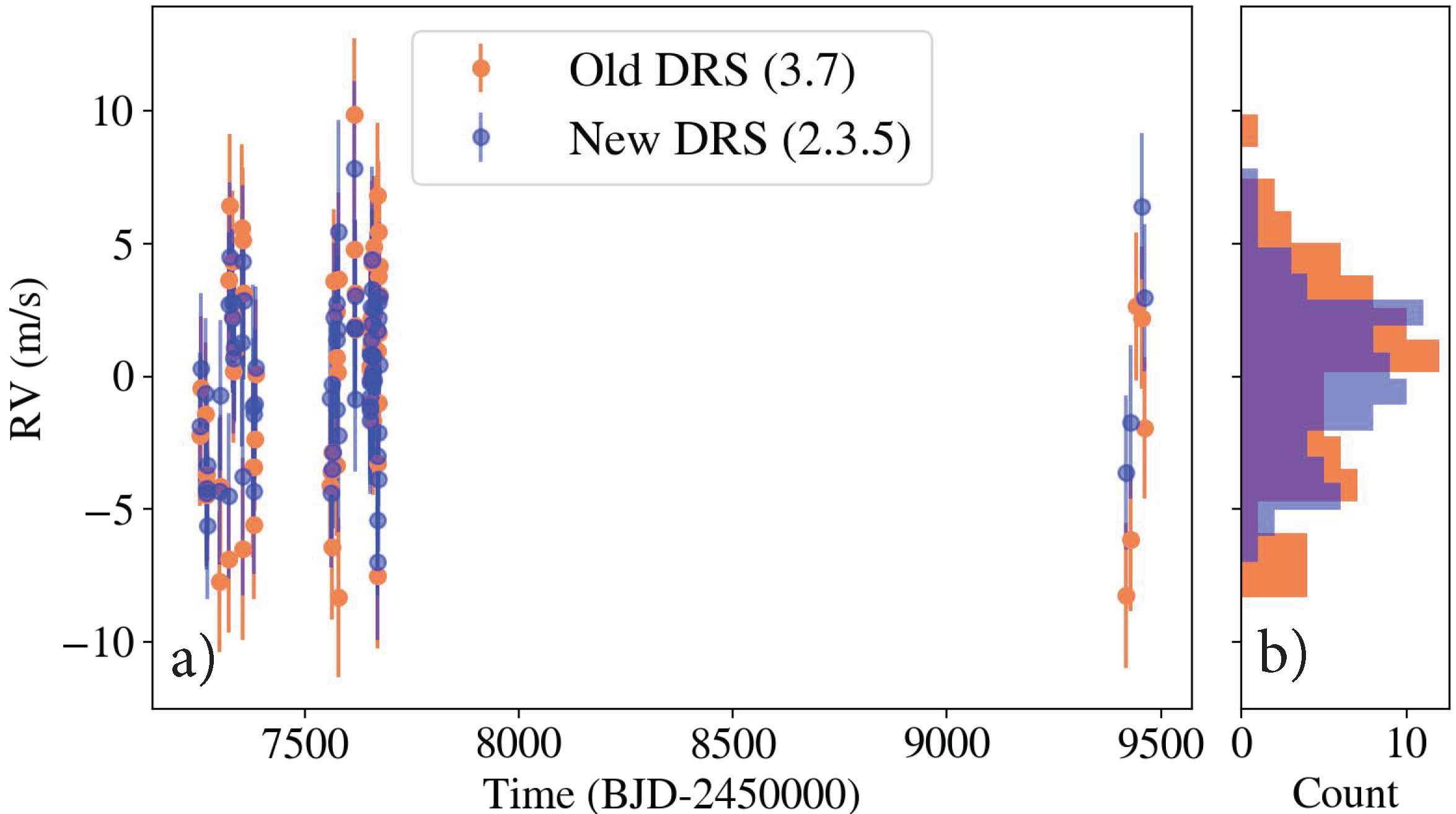}
    \caption{RVs from the HARPS-N old DRS (3.7) in blue and the new DRS (2.3.5) in orange. a) The RVs are plotted over time. There appears to be a slight upward trend toward the end of the \redit{time series}. b) Histogram of the old and new DRS. The new DRS has less scatter than the old DRS.  
    }
    \label{fig:timeseries}
\end{figure}

The 15 minute exposures of \thisstar\ yielded typical formal measurement uncertainties of 1.4 \ms. For a bright star like \thisstar\  \redit{($m_v = 8.24$)}, we expect this uncertainty to be a combination of instrumental systematics and the photon-limited uncertainty. We performed our analysis on \redit{an} older version of the HARPS-N pipeline (DRS 3.7) and the latest version of the pipeline (DRS 2.3.5) \redit{\citep{Dumusque2021}} and found significantly larger scatter for the older DRS pipeline, which we discuss extensively in Section \ref{rv_analysis}. \redit{There were numerous improvements between th\redit{is} old\redit{er} and new DRS to correct for systematics and instrumental effects. In addition, in the new DRS, different stellar masks were used that focus on spectral lines that are less sensitive to stellar activity and to more closely match the stellar type of the targets. In the case of \thisstar (F7-type star), DRS 3.7 uses a K5-type binary mask and the DRS 2.3.5 uses a G9-type binary mask}. For both of these pipeline version\redit{s}, the data reduction was carried out by cross-correlating the observations with \redit{one of these} binary mask\redit{s}. \redit{For the DRS 3.7, this produces a 161 element cross-correlation function (CCF), whereas for the newer 2.3.5 DRS, the sampling was reduced and the pipelines yields a 49 element CCF.} The\redit{se} resulting CCF\redit{s are}  used as an input representation for our \calm\ method \bedit{to track the activity-induced line shape changes in the spectra}. The DRS measures the radial velocities of each observation by fitting the CCF with a Gaussian function. However, a Gaussian is simple and symmetric and therefore unable to model the small CCF shape changes introduced by stellar activity. Thus, these radial velocities will include contributions from both Keplerian shifts and stellar activity signals. The full HARPS-N RV \redit{time series} from the new DRS is shown in Figure \ref{fig:timeseries}.

\subsection{Photometry}
\subsubsection{K2}
\begin{figure*}
    \centering
    \includegraphics[width=0.70\linewidth]{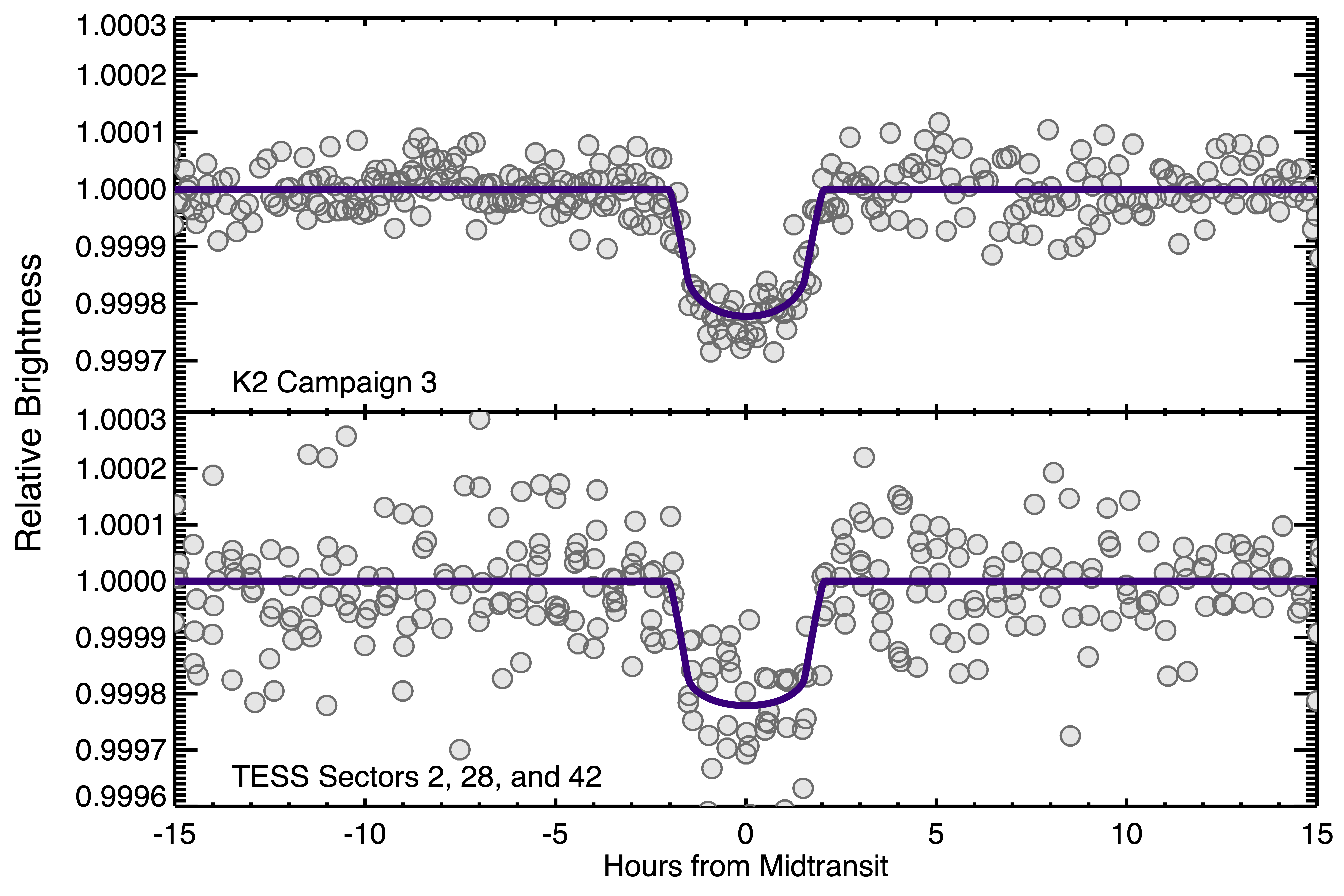}
    \caption{\Ktwo\ (\textit{top}) and \TESS\ (\textit{bottom}) light curves \redit{phase-folded on the period and centered at mid-transit from \Ktwo\ Campaign 3 and \TESS\ Sectors 2, 28, 42 . The purple lines shows the best fitting transit models.}}
    \label{fig:transits}
\end{figure*}

\thisplanet\ was observed by \Ktwo, \textit{Kepler}'s extended mission \redit{which} achieved similar precision to the original Kepler mission after applying systematic corrections \citep{2016ApJS..222...14V}. \thisstar\ was observed during K2 Campaign 3 from 2014 November 17 to 2015 January 23 in \redit{\Ktwo}'s long cadence mode (29.4 minute exposure times). We used the light curve produced by \citet{2018AJ....155..136M}. In brief, they extracted the light curve for \thisstar\ from the target pixel files, which were accessed through the Mikulski Archive for Space Telescopes (MAST)\footnote{mast.stsci.edu/portal/Mashup/Clients/Mast/Portal.html}. The light curves were reprocessed to simultaneously fit the known planet transit, remove mechanical \textit{K2} systematics, and fit stellar variability by using the methods described in \citet{2014PASP..126..948V} and \citet{2016ApJS..222...14V}. Due to the unusual brightness of the host star, \citet{2018AJ....155..136M} used larger photometric apertures than their standard analysis for this star. These apertures did not require dilution corrections because  \thisstar\ is sufficiently isolated from other stars or background objects \citep{2018AJ....156..182L}. To remove high-frequency variability, we divided the simultaneous-fit light curve by the best-fit spline component. The \Ktwo\ phase-folded transit light curves are shown in the top panel of Figure \ref{fig:transits}. For computational efficiency, we trimmed the light curve away from transit for our global fit, and only included a baseline of one transit duration on either side of the full transit (\bedit{See} Section \ref{transit_analysis}).

\begin{figure*}  \includegraphics[width=1.0\textwidth]{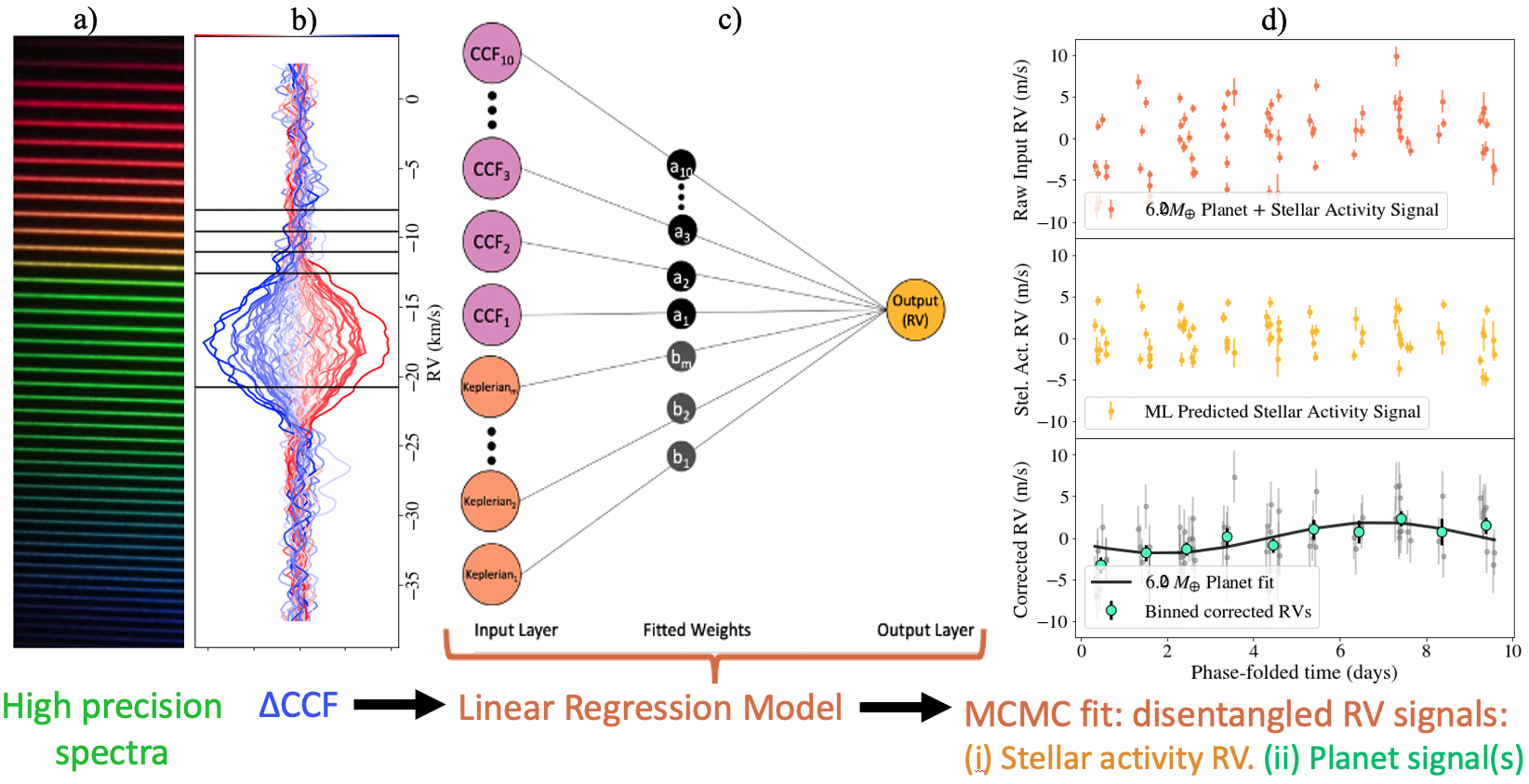}
    \caption{\beditr{Schematic of our \calm\ pipeline}. We use the changing shape of spectral features (measured by the change in the cross-correlation function or CCF) to predict and remove stellar activity signals. We use a linear regression model with \bedit{5-25} indexes from the CCF residuals used as inputs. We also simultaneously fit Keplerians along with the activity corrections to predict the overall RV signal. \bedit{The data shown in (b) and (d) are $\Delta$CCFs and the RVs of \thisstar\ respectively for the old DRS (3.7).}}
    \label{fig:simplifiedEPRV_pipeline}
\end{figure*}

\subsubsection{TESS}
\thisstar\ was pre-selected for two-minute cadence observations by \bedit{the} \TESS\ \bedit{mission} \citep{2015JATIS...1a4003R} and observed by Camera 1 during \TESS\ Sector 2 from 2018 August 22 to September 20, Sector 28 from 2020 July 31 to 2020 August 25, and Sector 42, from 2021 August 20 to 2021 September 16. The \TESS\ light curve was generated by NASA’s Science Processing Operations Center (SPOC) pipeline using the \texttt{Lightkurve} package \citep{2018ascl.soft12013L}. SPOC received raw data and processed the images to remove systematic erros and extract photometry \citep{2016SPIE.9913E..3EJ}. Using the SPOC Transiting Planet Search \citep{2002ApJ...575..493J}, the light curves are searched for transit events. \redit{The Pre-search Data Conditioned Simple Aperture Photometry (PDCSAP) flux uses the optimal \TESS\ aperture to extract the flux, corrects for systematics using PDC, and produces the light curves \citep{Stumpe2012, Stumpe2014, Smith2012}.} From these light curves we remove low frequency trends that correspond to astrophysical variability \redit{and remove remaining systematics based on the out-of-transit photometry} using \redit{\texttt{keplerspline}}\footnote{\redit{\url{https://github.com/avanderburg/keplersplinev2}}} \citep{Vanderburg2014, Shallue2018}.  The \TESS\ \beditr{phase-folded} transit light curves are shown in the bottom panel of Figure \ref{fig:transits}.

\subsubsection{Spitzer}
To perform follow up observations on the \textit{K2} transits, one transit of \thisplanet\ was observed with the \textit{Spitzer} InfraRed Array Camera \citep[IRAC;][]{Fazio2004} on March 1st 2016 as part of \redit{a} larger \textit{K2} follow-up program (Program ID: 11026; PI Michael Werner). The observations lasted for 10.16 hours and were taken with 0.4 second exposures using channel 2 on IRAC (2-4 $\mu$m). \beditrr{The observations are further described in detail in \citet{Duck2021} who found that the complete egress \redit{wa}s missing due to \redit{the} uncertainty in scheduling of the observations. Nonetheless, w}e reduced the data using BiLinearly
Interpolated Subpixel Sensitivity (BLISS) mapping \citep{Stevenson2012} and Pixel Level Decorrelation \citep{Deming2015, Garhart2020} and saw a feature consistent with a \beditrr{partial} transit, but with low enough SNR that it was not constraining to our model. We therefore exclude the \textit{Spizter} observations from our \redit{final} analysis.

\section{RV Analysis}\label{rv_analysis}
We performed an extensive series of analyses to determine the mass of \thisplanet. The RV \redit{time series} is shown in Figure \ref{fig:timeseries} where there appears to be a slight long-term upward trend in the RVs that may be the result of long-term activity cycles.  For our activity analyses, we first performed a simple RV analysis where we fit\redit{ted} a Keplerian signal and d\redit{id} not model the stellar variability. \redit{We ran Markov Chain Monte Carlo (MCMC) to sample the parameter space and model the Keplerian using  \texttt{radvel}'s \texttt{kepler.rv\_drive} function\footnote{We note that we accounted for the way that \texttt{radvel} used $\omega_p$ rather than $\omega_*$ in its implementation \citep{Householder2022} and only report $\omega_*$ in this paper.} \redit{\citep{Fulton2018}}. We use the differential evolution MCMC package \texttt{edmcmc\footnote{https://github.com/avanderburg/edmcmc}} \citep{andrew_vanderburg_2021_5599854}, which is an implementation of MCMC that incorporates differential evolution \citep[DE;][]{TerBraak2006} to solve the problem of choosing an appropriate scale and orientation for the jumping distribution, enabling a significantly faster fitting process. We computed 5,000 MCMC chains, discard the first 25\% as burn-in, and required that our parameters converged using the Gelman-Rubin statistic \citep{1992Gelman}, where our Gelman-Rubin threshold < 1.01. In this analysis, we observe\redit{d} that there is a large difference in RV scatter\footnote{Here we report the remaining RV scatter after removing our modeled Keplerian signal. In this way, we report only the unmodeled variation in the RVs as further detailed in Section \ref{scatter_metrics}.} between the two DRS versions (3.95 \ms\ and 3.01 \ms\ for the old and new DRS, respectively) and that the uncertainties on the estimated semi-amplitude (K) for the Keplerian signal are larger for the old DRS ($2.0_{-0.7}^{+0.7}$ \ms) than the new DRS ($2.0_{-0.5}^{+0.5}$ \ms). The expected scatter due to \redit{instrument systematics (photon noise, drift noise, and wavelenegth calibration errors)} for the \redit{both} DRS \redit{versions} is $\sim$ 1 \ms, which is} significantly less than \redit{the scatter that} we measure\redit{d}. \redit{For this chromospherically active star \thisstar\ ($\textrm{log R}_{\textrm{HK}}^{'} = -4.66$), stellar variability could be causing a part of this} excess scatter (and the \redit{relatively} large uncertainty in K). Thus, we sought to develop new methods that could model this stellar variability by tracing shape changes in the CCF. \redit{For each of our RV analyses, we performed an independent analysis using only the 3.7 DRS and an independent analysis using only the 2.3.5 DRS observations. We included both DRS versions in our RV analyses since we found interesting differences in the effectiveness of our activity mitigation methods for the two versions as discussed extensively in Section \ref{mcmc_results}.}

\subsection{\calm\ Method}
In order to account for the stellar variability we observe in the RV measurements of \thisstar, we developed a simplified machine learning model \calm\ that allows us to separate the stellar activity from the \beditr{D}oppler reflex motion. This simplified method is inspired by the previous implementation of a neural network-based stellar activity mitigation method for solar observations using line shape changes to predict stellar activity signals \citep{deBeurs2022b}. To extend this approach to extrasolar observations, a new simplified method was required for two primary reasons:
\begin{enumerate}
    \item RV datasets for \redit{other} stars rarely have as many observations as the solar dataset from the HARPS-N Solar Telescope used in \citet{deBeurs2022b}, which consisted of $\sim$600 days of observations.
    \item We lack a firm ``ground-truth'' for the stellar activity signals of \redit{other} stars. For the Sun, we can easily remove the RV contributions of the known planets in the Solar System, but this is not as straighforward for other stars. \redit{This is because there may always be unmodeled planetary signals that fall below the detection threshold. These unmodeled planetary signals may add noise to our estimates of the stellar variability signal of a given star.}
\end{enumerate}

To overcome these \bedit{challenges}, we made two modifications to our method. First, we greatly simplified the machine learning models for use on stars other than the Sun. Instead of using complex and flexible models like our convolutional neural networks, we use a linear \beditr{regression model} with a \beditrr{significantly smaller} input representation. \redit{In particular, the input representation consists of} 5\redit{-25} CCF activity indicators\footnote{\redit{We tested a range of input representation sizes as discussed in Section \ref{sign_indexes} and ultimately find that 5 CCF activity indicators to balances the risk of overfitting and underfitting to the data for the 2.3.5 DRS. For the 3.7 DRS, we find that 25 CCF indexes balances these concerns.}} instead of the entire CCF array\redit{, which contains 49 elements for the 2.3.5 DRS and 161 elements for the 3.7 DRS}. Essentially\beditr{,} we have sacrificed some fidelity in our stellar activity corrections in exchange for immediate applicability to a much larger sample of stars. Eventually\beditr{,} it may be possible to train a more complex model on the entire ensemble of stars observed, but our simplified approach is a first step \bedit{towards this goal}. 

The second modification we made to our technique for use on stars other than the Sun \bedit{involves fitting for planet signals simultaneously.} We cannot \textit{a priori} remove all planetary signals from the dataset (since \redit{the amplitude and period of such signals are unknown and/or they} may be hidden by stellar activity). \bedit{Thus,} we must fit any remaining planetary signals \textit{simultaneously} with the stellar activity correction. Although this increases the computational difficulty of our method (because we must test many different possible planetary signals to discover new planets), the increase in computational difficulty is offset by the simplicity of our linear \bedit{regression} model. An overview of our simplified approach is shown in Figure \ref{fig:simplifiedEPRV_pipeline}.


\beditr{We can describe this simplified approach as} simultaneously fitting for $n$ Keplerian signal\beditr{(s)} and a linear regression model with \beditr{\textit{m}} CCF activity indicators for stellar \bedit{activity}. \beditr{The number of CCF activity indicators \textit{m} is determined by finding a balance between optimizing goodness-of-fit and preventing overfitting. In equation form}, our model is given as:

\begin{equation}\label{LR_model}
\begin{split}
        RV = \textcolor{blue}{w_1 \cdot CCF_1 + w_2 \cdot CCF_2 +  ... + w_{m} \cdot CCF_{m}} \\
        + b_1 \cdot Kepler_1 + b_2 \cdot Kepler_2 + ... + b_n \cdot Kepler_n \\
        = \textcolor{blue}{RV_{\rm activity}} + RV_{\rm Keplerians}
\end{split}
\end{equation}

where $CCF_m$ is the value of the CCF at \redit{index} $m$, $w_m$ is a weight parameter for $CCF_m$, $Kepler_n$ is the $n_{th}$ fitted Keplerian, and $b_n$ is this Keplerian's weight parameter. \redit{This Keplerian weight parameter $b_n$ is equivalent to a planet's semi-amplitude ($K$), but parametrizing the model in this way allows us to do a linear least squares regression fit where we can simultaneously model planetary and stellar variability contributions to the overall RV signal.}

This section describes our new \calm\ method in detail. Section \ref{ccf_inputs} describes the pre-processing we must apply to run our \calm\ method; in particular, we first must prepare the CCF input representation for our model by centering the data and choosing the most information-rich subset of the CCF to feed into the model. Before running our models, we check several diagnostics to prevent overfitting, which are described in Section \ref{no_more_overfitting}. Next, we sample the parameter space using a differential evolution-based MCMC method described in Section \ref{DE_MCMC}. Finally, we evaluate our models and compare them with several traditional activity indicators using a new scatter metric described in Section \ref{scatter_metrics} and describe our RV analysis results in Section \ref{mcmc_results}.

\begin{figure}
\centering
    \includegraphics[width=0.9\linewidth]{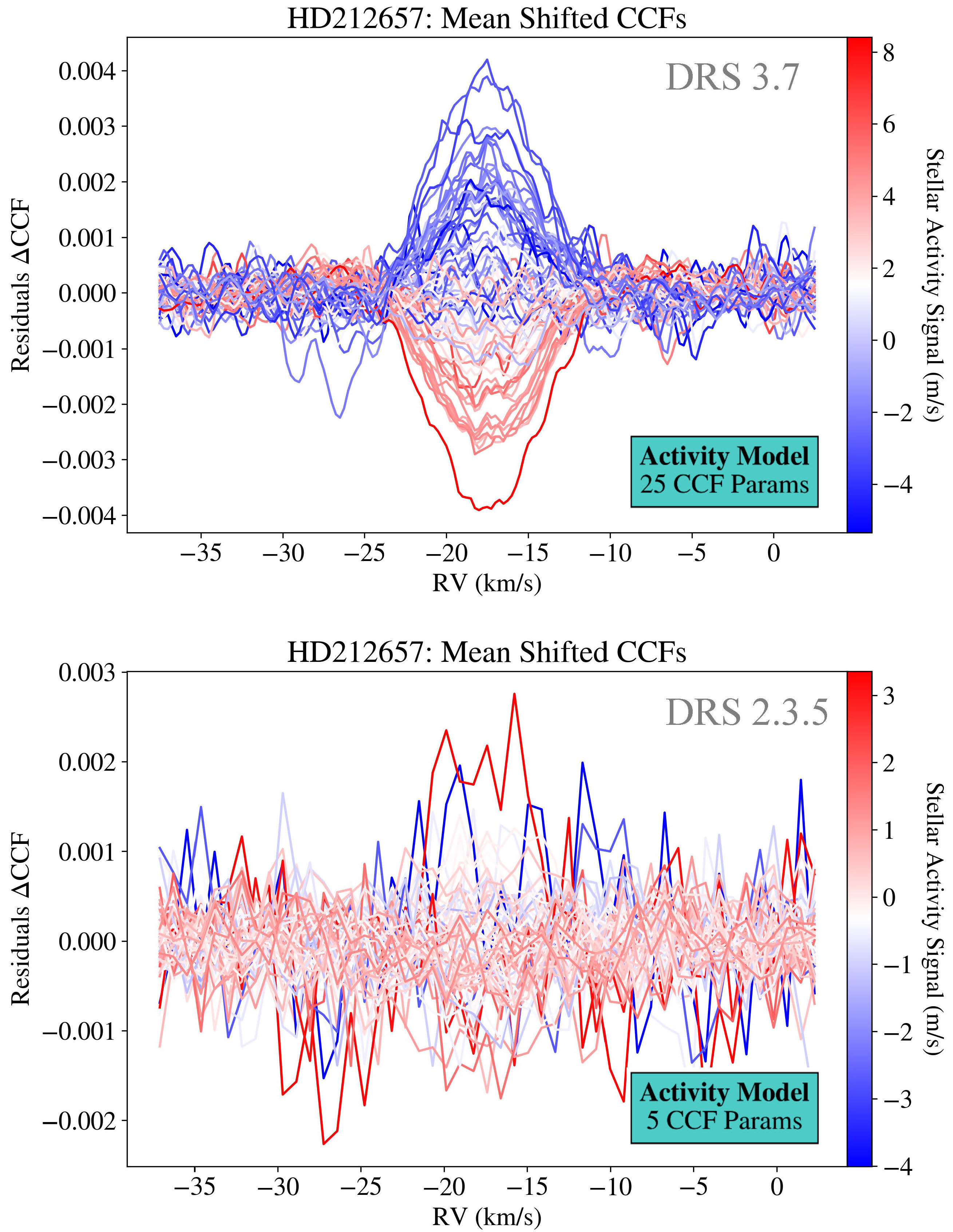}
    \caption{Computed residual CCFs ($\Delta$CCFs) for both the old\redit{er} (3.7) and new\redit{er} DRS (2.3.5). $\Delta$CCFs are computed by subtracting the mean CCF and highlight differences in features between CCFs. The model predicted stellar activity contribution to the RVs is indicated by its color (red = redshifted, blue = blueshifted). }
    \label{fig:residccfs}
\end{figure}

\subsection{Preparing the CCF Input Representation}\label{ccf_inputs}
\subsubsection{Computing and Centering our $\Delta$CCFs}\label{center_ccfs}
For our \calm\ model, we have to pre-process our data products into a uniform format which makes capturing \bedit{and learning} features in our data easier. We design the input representation such that the model becomes sensitive to shape changes due to stellar activity, not translational shifts caused by true \beditr{D}oppler shifts. Simply put, we want the \calm\ model  to find the difference between a Gaussian fit to the CCF and the true center-of-mass velocity shift. In this way, our model will be able to disentangle stellar activity signals from true Keplerian shifts.






To frame the problem in this way, we shift every CCF such that the velocity predicted by the Gaussian fit to the CCF is at the median velocity across all observations. \redit{We note that we took precautions to prevent improperly centering the CCFs in a way that planetary reflex motion remains in the input data. This could cause overfitting and we further describe our precautionary steps in Section \ref{no_more_overfitting}.} Next, we subtract a median CCF from each CCF to compute the residual CCF\footnote{We note that subtracting this median is purely for visualization purposes and not required for the regression algorithm.} ($\Delta$CCF), which are visualized in Figure \ref{fig:residccfs}, where they are color-coded based on how blueshifted or redshifted the RV signals are. There exists a \beditrr{strong signal} in shape changes based on these RV signals \beditrr{for the old DRS and a hint of a dependence for the new DRS}. \redit{This difference may be explained by the new DRS using spectral lines less sensitive to stellar variability and the difference in stellar type mask used.} \redit{Overall, we want } our \calm\ model  \redit{to learn how these shape changes relate to the stellar activity contributions to RVs so that we can model and mitigate the stellar variability using the $\Delta$CCFs}. 

After computing the $\Delta$CCFs, we normalize the observations to ensure the scale of variations across each input parameter is approximately equal. A subset of indexes across these normalized $\Delta$CCFs are then fed into the \calm\ model . We choose to only include a subset of indexes (\beditr{5 to 25}\footnote{The exact number is determined by optimizing the fit while preventing overfitting as described in Section \ref{no_more_overfitting}} indexes) rather than the entire normalized $\Delta$CCFs (\bedit{49 indexes or 160 indexes for DRS 2.3.5 and DRS 3.7 respectively}) because we only have $\sim$80 RV observations and the entire normalized $\Delta$CCFs would provide \beditr{too many free parameters}. We describe how we select which CCF \redit{indexes} to use in Section \ref{sign_indexes}. 

\subsubsection{Selecting the most significant indexes of the $\Delta$CCF}\label{sign_indexes} 
\begin{figure}
    \includegraphics[width=1\linewidth]{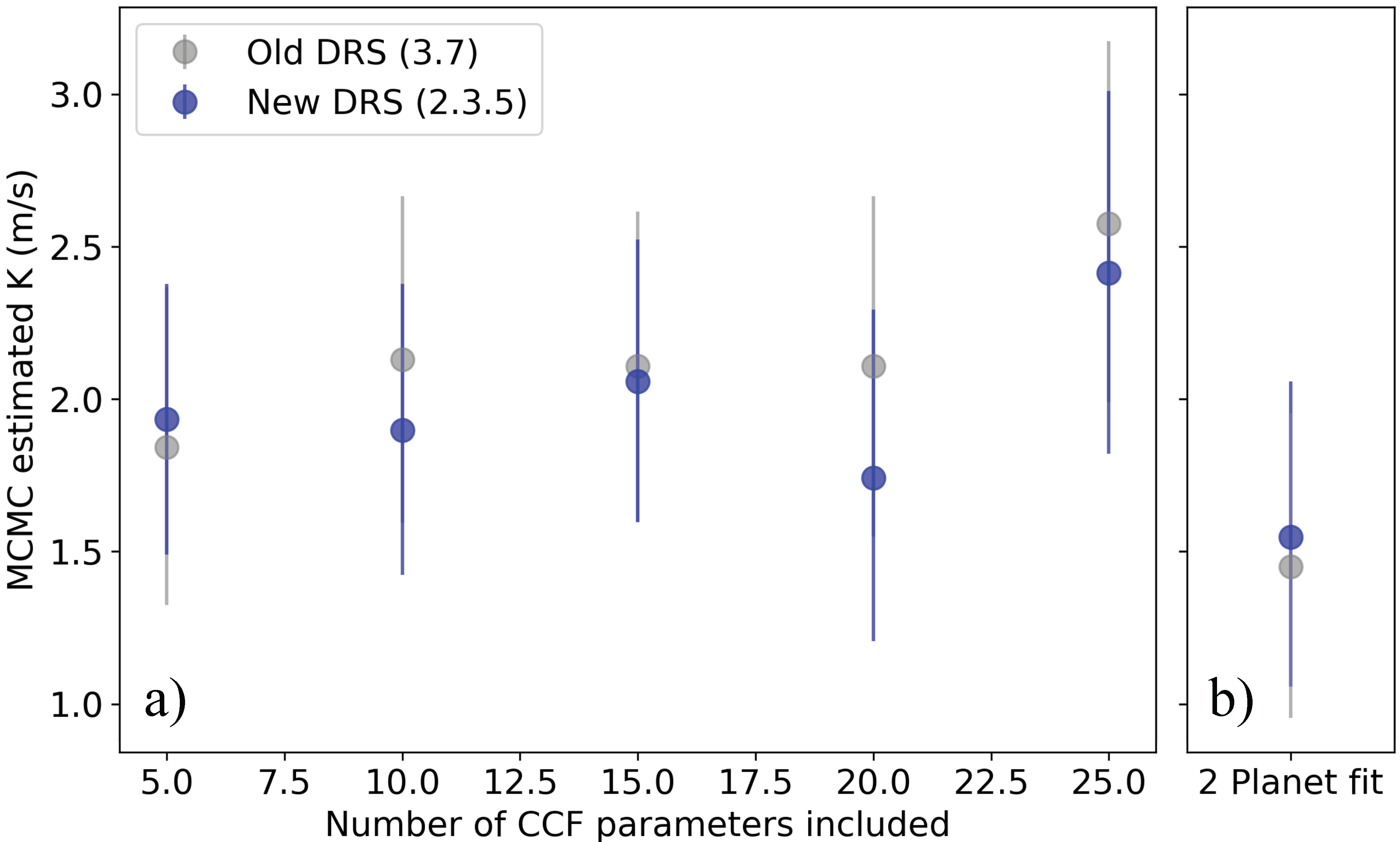}
    \caption{\bedit{Estimated semi-amplitude for various model configurations. (a) }Estimated semi-amplitude (K) as function of number CCF \redit{indexes} for both the old DRS (grey) and new DRS (blue). As the number of CCF \redit{indexes} included increases, there a slight positive trend in K. \bedit{(b) Estimated K for \thisplanet\ in a two-planet fit model as described in Section \ref{2nd_companion}.} Generally, the estimates of K agree within \redit{the} error\redit{s} for any of these model configurations.}
    \label{fig:K_consistent}
\end{figure}

To ensure that our number of free parameters is significantly less than our number of observations (74), we include only a subset of the CCF rather than the entire 49-element array of the CCF \footnote{the entire array of the CCF contains 161 values for the old DRS (3.7). This was adjusted to 49 in the new DRS (2.3.5) to correct for oversampling of the CCF.}. We choose the subset of the CCF by checking which indexes provide \beditr{the most} statistically significant contribution in Equation \ref{LR_model}. We check for statistical significance by sampling the parameter space with Differential Evolution MCMC (described in Section \ref{DE_MCMC}) and estimating uncertainties on the CCF weights ($w_1, w_2, ..., w_n$). \beditr{Commonly, when including $\geq 30 $ CCF \redit{indexes}, the \redit{information at these }indexes ha\redit{s} degeneracies which dilutes the significance, resulting in the highest significance indexes only having $1-2\sigma$ significance. This means that for these indexes the value of the weights for $CCF_i$ is one to two times its corresponding uncertainty.}  We then choose the 5-25 most statistically significant CCF \redit{indexes} for our final model. \beditr{This exact number depends on the DRS version as explained in detail in Section \ref{results_in_fourier_space}. When only including those ``significant indexes'' in our model, the significance increas\redit{es} to $3-5\sigma$. Generally, we find that the estimated semi-amplitude (K) of the Keplerian signal of \thisplanet\ is consistent for any of the number of CCF \redit{indexes} tested (Figure \ref{fig:K_consistent}).}


\subsection{Running our RV models with Differential Evolution  MCMC}\label{DE_MCMC}
Once we have chosen our CCF \redit{indexes} and the number of \redit{K}eplerians to include in our model, we sample our parameter space using \redit{differential evolution MCMC implemented in} \texttt{edmcmc\footnote{https://github.com/avanderburg/edmcmc}} \citep{andrew_vanderburg_2021_5599854}. We implemented Gaussian priors to the $P$ and  $t_{\textcolor{new_color}{c}}$ parameters based on the EXOFASTv2 transit fit results (Table 1)\redit{, which are discussed in Section \ref{transit_analysis}}. \redit{We note that without including priors on  $P$ and $t_c$, we recover the period and time of conjunction in the RVs. However, we include the priors in our analysis since they allow us to get the tightest constraints on the mass of \thisplanet.} We model the Keplerian signal of \thisplanet\ by using \texttt{radvel}'s \texttt{kepler.rv\_drive} function \redit{\citep{Fulton2018}}. After performing the MCMC fit, we determine convergence of the parameters by using the Gelman-Rubin statistic \citep{1992Gelman}, where our Gelman-Rubin threshold < 1.01. We computed 5,000 MCMC chains and discard the first 25\% as burn-in before producing our posterior samples. We compute the median and 68.3\% confidence intervals to derive our final parameter values. The results of our DE-MCMC analysis of the RVs are described in Section \ref{mcmc_results}.  

\begin{figure*}
    \includegraphics[width=0.95\textwidth]{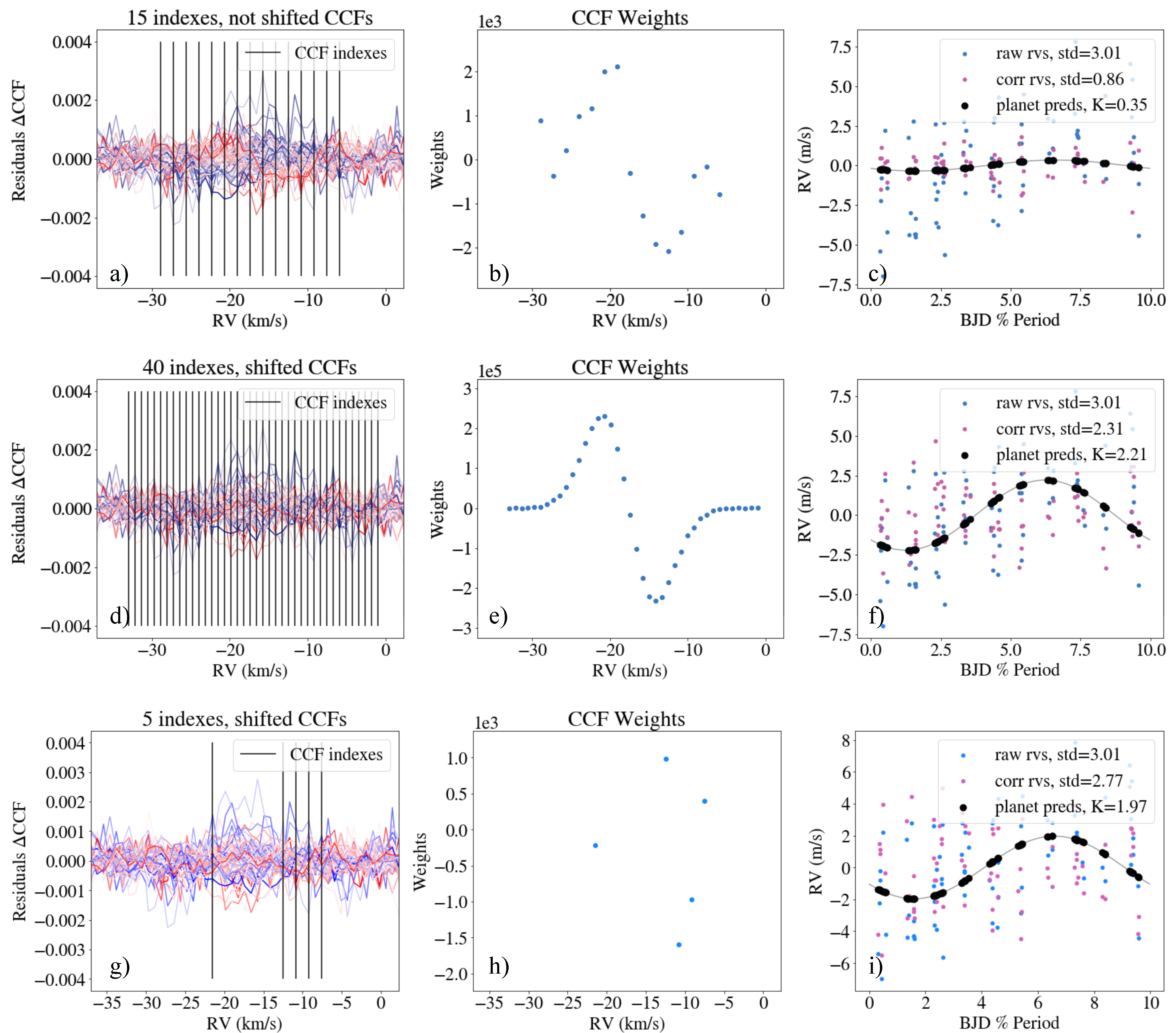}
    \caption{Overfitting diagnostics for stellar activity RV analysis using the new DRS (2.3.5) \bedit{HARPS-N $\Delta$CCFs}. \redit{The three columns in this Figure show the $\Delta$CCFs, the weight parameters corresponding to the $\Delta$CCFs, and the \redit{phase-folded} RVs for three scenarios respectively. For each of the Figures in the third column \rredit{(c, f, i)}, we provide a legend which includes the scatter of the radial velocities from the HARPS-N pipeline (raw rvs, std) in \ms, the scatter of the activity corrected radial velocities (corr rvs, std) in \ms, and the predicted semi-amplitude of the planet (planet preds, K) in \ms. The first two rows corresponds to two different overfitting concerns and the last row demonstrates a case where both of these concerns are addressed.}  In the first row \rredit{(a, b, c)}, we plot the $\Delta$CCFs that have not been shifted to be centered at the median velocity. Without centering the $\Delta$CCFs, the algorithm will be able to access translational shift \bedit{(i.e. \beditr{D}oppler shifts)} information and attribute potential planet signals to stellar activity signals\beditrr{, significantly attenuating their amplitude}. 
    In the second row\rredit{(d, e, f)}, we show a case where the $\Delta$CCFs are shifted but the number of indexes is too large\beditrr{. T}his results in overfitting as seen in the weights plot for the second row. In the third row\rredit{(g, h, i)}, we have only fed in the $\Delta$CCF \redit{indexes} that are considered significant (as described in Section \ref{sign_indexes}) and use the \bedit{properly} shifted $\Delta$CCFs. }
    \label{fig:overfitting diagnostics}
\end{figure*}

\subsection{Overfitting Diagnostics}\label{no_more_overfitting}

Overfitting is a significant problem in radial velocity analyses and the state-of the art methods that are widely used to model and predict stellar variability often do not agree within error. The severity of this problem is particularly evident in recent work by \citet{Zhao2022} where 11 teams were invited to use their 22 different methods on ultra precise EXPRES radial velocities of four stars that encompass various levels of activity. There is a concerning lack of agreement between these methods across these targets, which \redit{perhaps can be} attributed to some of the methods overfitting to noise in the observations. \redit{Recently, \citep{Blunt2023} demonstrated that Gaussian process regression, one of the most common and state-of-the-art methods of modeling stellar variability, can be susceptible to overfitting.}  We wanted to prevent \redit{overfitting} and \redit{thus} set out to develop various methods to \redit{mitigate the risk of overfitting}.

We developed several diagnostic plots as illustrated in Figure \ref{fig:overfitting diagnostics}. We developed these diagnostics to prevent two common types of \bedit{overfitting} concerns:
\begin{enumerate}
    \item Not properly centering CCFs such that plane\redit{t}ary reflex motion remains in the input data.
    \item Using too many CCFs indexes such that there are too many free parameters compared to number of observations
\end{enumerate}

\begin{figure*}
    \includegraphics[width=1.02\textwidth]{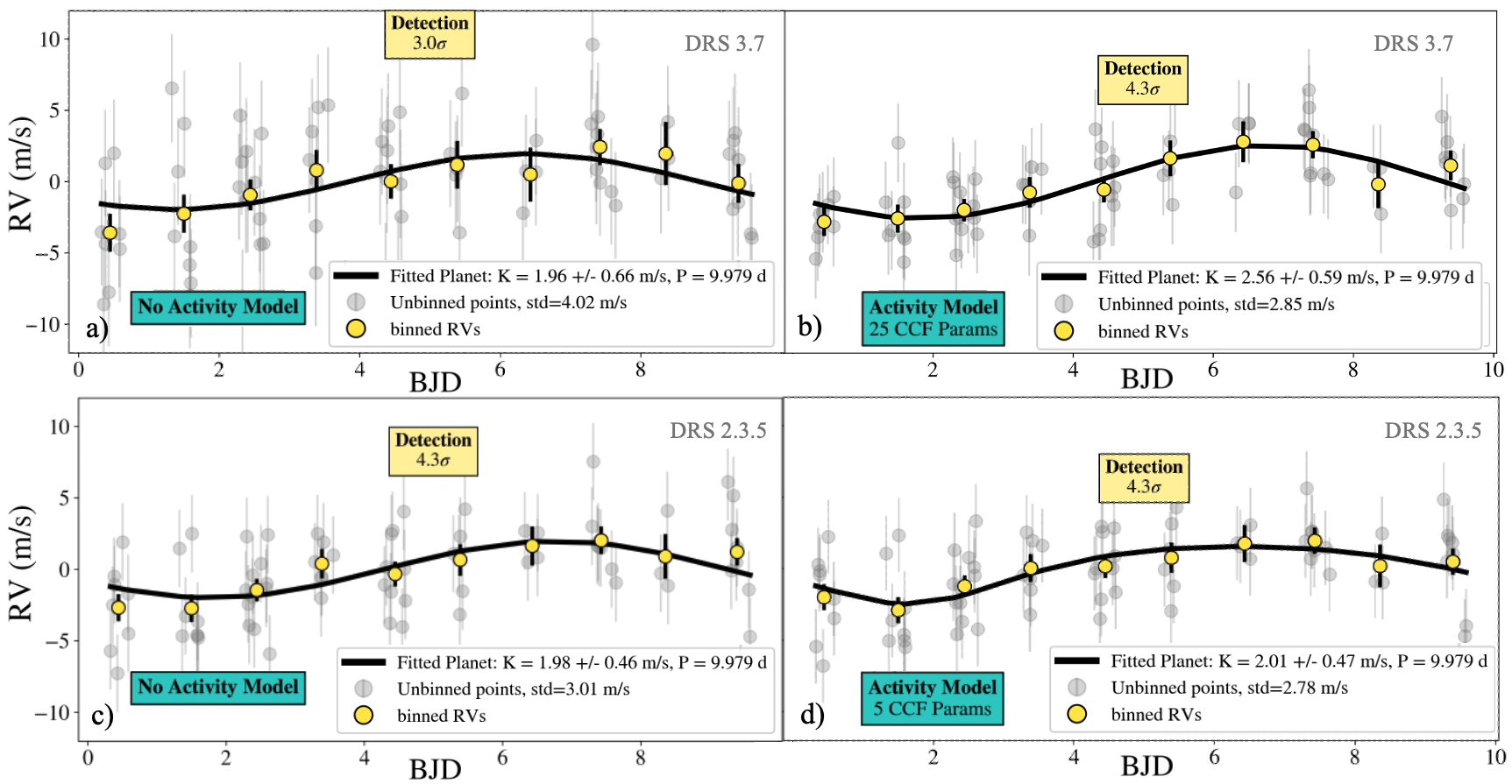}
    \caption{Phase-folded RVs before and after applying our activity correction model. From the first to second column, we compare the \redit{phase-folded} RVs  for the old (3.7) and new (2.3.5) DRS. (a) Before applying our linear regression activity correction, the RV scatter is $\sigma_{\rm raw}$ = \olddrsrmsRAW\ \ms\ and  we detect the semi-amplitude of the planet with 3.0$\sigma$ confidence in the old DRS (3.7) RVS. (b) When we simultaneously fit our \calm\ activity model based on shape changes in the CCF with a Keplerian, the detection significance increases to 4.3$\sigma$. c) For the new DRS, the raw scatter is much lower ($\sigma_{\rm raw}$ = \newdrsrmsRAW\ \ms) and without applying an activity model, we get a 4.3$\sigma$ detection. d) With an activity model, the raw scatter decreases to $\sigma_{\rm raw}$ =  \newdrssigmaUS\ \ms and the detection significance remains the same at 4.3$\sigma$. For a description of the scatter metrics, $\sigma_{\rm raw}$ and $\sigma_{\rm us}$, see Section \ref{scatter_metrics}.}
    \label{fig:hd212657}
\end{figure*}

\noindent To address the first concern, we plot the $\Delta$CCFs in the first column \bedit{of Figure \ref{fig:overfitting diagnostics}} \bedit{to} check whether all translation shifts (i.e. planetary reflex motion) were removed and the CCFs were properly centered. To illustrate \bedit{how} this problem \bedit{appears}, we intentionally did not center the $\Delta$CCFs in Figure \ref{fig:overfitting diagnostics}a. We thus observe a sinusoidal-like pattern for the red $\Delta$CCFs which correspond to more redshifted RVs and the opposite pattern for more blue $\Delta$CCFs. This is a tell-tale sign of a translation shift still remaining within the $\Delta$CCFs. We can understand this shape by reminding ourselves of the definition of a derivative where we define the median CCF as $f(x)$ and a redshifted CCF as $f(x-h)$ and thus can write
\begin{equation}
        f'(x) = \lim_{h \to 0} \frac{f(x-h) - f(x)}{h}
\end{equation}
Thus, the shape expected from the difference between the redshifted CCF and the median CCF is simply the derivative of a Gaussian. This is illustrated in Figure \ref{fig:ccf_shapes} where we simulated a few Gaussian curves \bedit{and added white Gaussian noise. We} show the expected residuals for improperly centered and properly centered CCFs in Figures \ref{fig:ccf_shapes}a,b and \ref{fig:ccf_shapes}c,d respectively. When $\Delta$CCFs are properly shifted to the center, this pattern disappears as seen in both the simulated example in Figure \ref{fig:ccf_shapes}c,d and in the real data in Figure \ref{fig:overfitting diagnostics}d,g. 

To address the second concern, we created diagnostic plots that include the weights \bedit{for the $\Delta$CCFs} as seen in the second column of Figure \ref{fig:overfitting diagnostics}. When too many CCF weights are included, we tend to see structure in these weights plots that is caused by overfitting \bedit{as one weight is compensating for its neighboring weight}. We illustrate this in the second row of Figure \ref{fig:overfitting diagnostics} where we show a case where the $\Delta$CCFs are shifted properly but the number of indexes is too large. Although this result in quite a low RMS of 2.3 \ms, the large number of free parameters results in overfitting on the noise in the data. We note that even when this overfitting occurs, the planetary signal remains preserved with a semi-amplitude of K $\sim$ 2.2 \ms.

Finally, we show an example in the third row of Figure \ref{fig:overfitting diagnostics} where both the $\Delta$CCFs are properly centered and the number of indexes (5) is small enough to not result in overfitting. Both the derivative-of-a-Gaussian pattern in the $\Delta$CCFs and the structure in the weights disappear.

\subsection{Introducing a new performance metric\beditr{: unexplained scatter}}\label{scatter_metrics}
To measure the original scatter in the data before applying any activity correction, we compute the raw scatter, $\sigma_{\rm raw}$, which is defined as the sample mean standard deviation of the RVs minus the expected \beditr{measurement error} such that:
\begin{equation}
    \sigma_{\rm raw} = \sqrt{SD(RV_{\rm HARPS})\rredit{^2}-\sigma_{\rm \beditr{measurement}}^2}
\end{equation}
where $\sigma_{\rm \beditr{measurement}}$ is the median estimated RV errors from the DRS which includes photon noise, drift noise, and errors in the wavelength calibration \citep{Dumusque2021}. Defining the scatter \redit{in} this way assumes that we can approximate the source of RV variations as Gaussian noise and that they can be approximated as independent scatter. Both of these assumptions are not true but this provides a sufficient approximation of the noise. \bedit{We then evaluate the reduction in RV scatter using} two scatter metrics: the activity corrected scatter ($\sigma_{\rm ac}$) and the unexplained scatter ($\sigma_{\rm us}$). We define $\sigma_{\rm ac}$ as
\begin{equation}
    \sigma_{\rm ac} = \sqrt{(SD(RV_{\rm HARPS}-RV_{\rm SA}))^2-\sigma_{\rm \redit{measurement}}^2}
\end{equation}
where SD is the sample standard deviation, $RV_{\rm HARPS}$ are the raw radial velocities from the HARPS DRS pipeline, $RV_{\rm SA}$ are the stellar activity \bedit{contributions to the} radial velocities \redit{as predicted by the given activity model}. This type of scatter metric \footnote{\bedit{Sometimes the convention is to not subtract the measurement errors from the instrument ($\sigma_{\rm measurement}$}) in calculating this scatter metric.} is sometimes used to evaluate and compare stellar activity models, but this metric does not account for the scatter that can be caused by planet signal(s). Focusing on just reducing this scatter metric could unintentionally incentivize removing planet signals rather than preserving them with these activity correction methods. To partially address this concern, we introduce a metric that subtracts modeled planet signals so we primarily evaluate the activity model correction method on unexplained scatter rather than any scatter. We define the unexplained scatter ($\sigma_{\rm us}$) as
\begin{equation}
    \sigma_{\rm us} = \sqrt{(SD(RV_{\rm HARPS}-RV_{\rm SA}-RV_{\rm planet}))^2-\sigma_{\rm \redit{measurement}}^2}
\end{equation}
where $RV_{\rm planet}$ is the contribution of modeled planets to the RV measurements. \bedit{We note that there can evidently still remain un-modeled planet signals so this scatter metric does not fully address the concern of incentivizing removal of planet signals to reduce RMS. However, this can be a step towards finding better metrics for evaluating our RV models.}

\begin{figure*}
    \centering
    \includegraphics[width=0.9\textwidth]{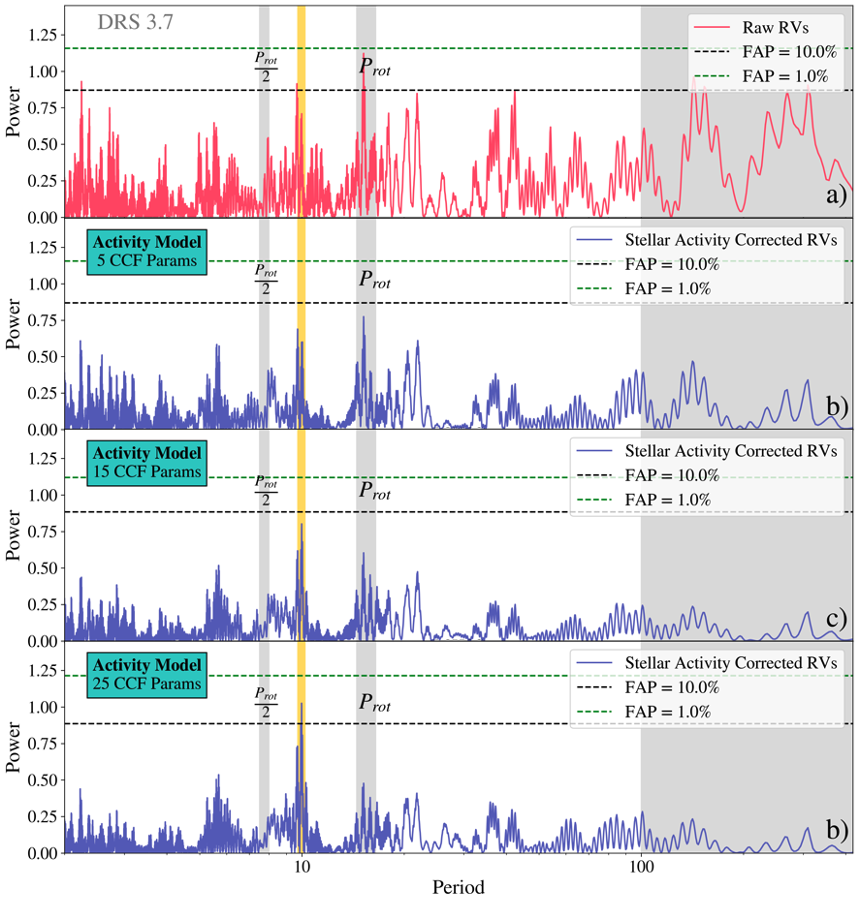}
    \caption{Old DRS (3.7) Periodograms as a function of number of CCF \redit{indexes} used in the stellar activity correction model. HARPS-N Raw (a) and Corrected (b) RVs in Fourier space for 5 CCF \redit{indexes}. \rredit{In each panel, the 1.0\% and 10.0\% false alarm probabilities (FAP) computed using the bootstrap method are indicated with green and black dotted lines, respectively.} The long-term activity signals in panel (a) decrease in magnitude after applying the activity correction in panel (b). The suspected rotation period and half the rotation period are indicated in grey. The planet period is indicated in yellow. As we increase the number of included CCF \redit{indexes} to N=15 (c) and N=25 (d), the peak corresponding to the star's rotation period ($P_{\rm rot}$) decreases significantly in magnitude after applying the stellar activity corrections and a planet signal emerges at \litperiod\ days.}
    \label{fig:old_drs_correctedperiodogram}
\end{figure*}

\begin{figure*}
    \centering
    \includegraphics[width=0.9\textwidth]{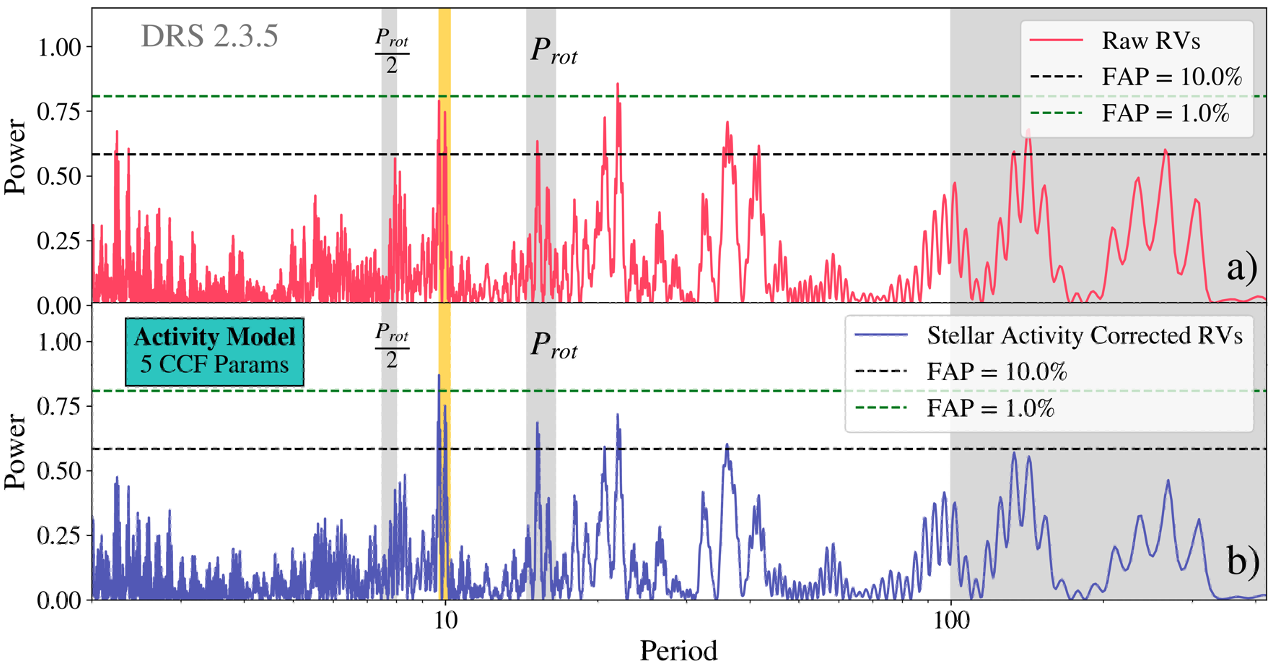}
    \caption{New DRS (2.3.5) Periodograms for 5 CCF \redit{indexes}. HARPS-N Raw (a) and Corrected (b) RVs in Fourier space for 5 CCF \redit{indexes}. \rredit{The 1.0\% and 10.0\% false alarm probabilities (FAP) computed using the bootstrap method are indicated with green and black dotted lines, respectively.} The long-term activity signals in panel (a) decrease in magnitude after applying the activity correction in panel (b). The suspected rotation period and half the rotation period are indicated in grey. The planet period is indicated in yellow. From (a) to (b), the peak corresponding to a planet signal at \litperiod\ days increases slightly in magnitude.}
    \label{fig:new_drs_correctedperiodogram}
\end{figure*}

\subsection{MCMC Results}\label{mcmc_results}

For both the new and old DRS, the MCMC successfully converged to a solution for the stellar activity model and the planet parameters. 

For the old DRS (3.7), we found that \calm\ stellar activity mitigation method yields a significant improvement in both the RV scatter (from $\sigma_{\rm raw}$ = \olddrsrmsRAW\ \ms to $\sigma_{\rm us}$ = \olddrssigmaUS \ms) and the precision with which we recover the mass of the transiting planet (improving a 3.0$\sigma$ significance detection into a 4.3$\sigma$ significance detection; see Figure \ref{fig:hd212657}a, b). With our MCMC fit, we derive a mass of
\calmmass \me\ and find that the orbital period and phase of the detection agrees with the known \litperiod\ day period from transit observations.

For the new DRS (2.3.5), we find that the \calm\ method does improve the scatter in the RVs from from $\sigma_{\rm raw}$ = \newdrsrmsRAW\ \ms\ to $\sigma_{\rm us}$ = \newdrssigmaUS \ms. However, the level of stellar activity is already sufficiently low in the new DRS that even without an activity model, we can detect the planet signal at 4.3$\sigma$ significance (Figure \ref{fig:hd212657}c,d). When applying our \calm\ framework, the detection significance remains at 4.3$\sigma$. Although this does not provide a notable improvement, this result does demonstrate that our method effectively preserves planetary \beditr{D}oppler signals rather than removing them to achieve a lower RV scatter. We note that after applying our activity mitigation method, the final unexplained scatter in the new DRS ($\sigma_{\rm us}$ = \newdrssigmaUS \ms) is marginally higher than in the old DRS ($\sigma_{\rm us}$ = \olddrssigmaUS \ms). This may be due to a difference in line lists between the old and new DRS where less activity sensitive lines were used to derive the CCFs and RVs in the new DRS. Although this may result in cleaner RVs, this may also make it more challenging to remove remaining stellar activity signals with the CCFs in the new DRS. In the future, using custom line lists that are especially sensitive to stellar activity to compute custom CCFs may allow us to mitigate the stellar activity more effectively using \calm.

\begin{figure*}
    \includegraphics[width=1.03\textwidth]{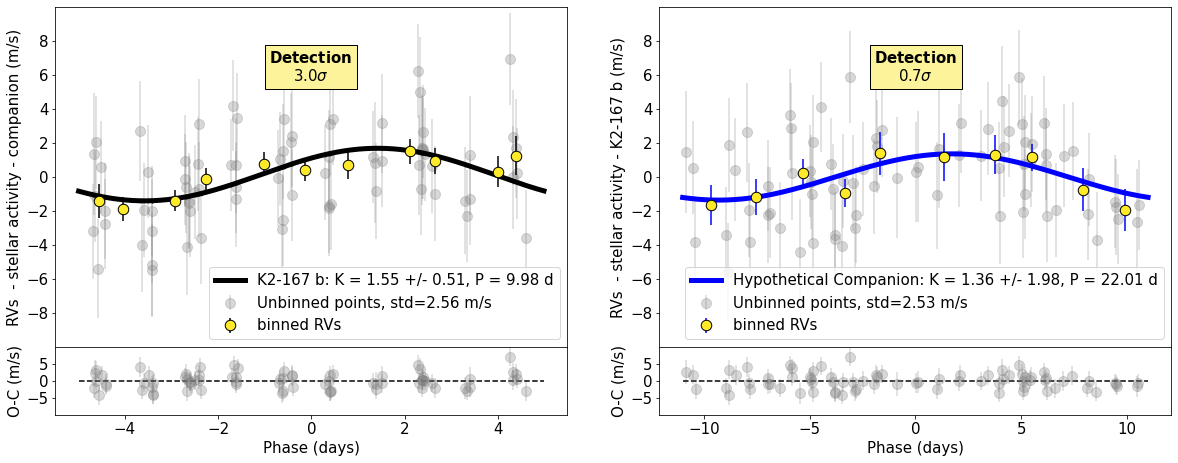}
    \caption{Phasefolded two planet fit. a) The RV model for \thisplanet\ is plotted in black and the binned RVs are plotted in yellow. The data is \redit{phase-folded} on \thisplanet's period of  \litperiod\ days. b) RV model for a hypothethical outer companion is plotted in blue and the binned RVs are plotted in yellow. The data is \redit{phase-folded} on the fitted period of 22.01 days.}
    \label{fig:two_planets_phasefold}
\end{figure*}

\subsubsection{Results in Fourier Space: Activity Signal Peaks Decrease and Planetary Signal Emerges}\label{results_in_fourier_space}

For both the old DRS and new DRS, we investigate the behavior of raw and stellar activity corrected HARPS RVs in the Fourier domain to \bedit{investigate} which \bedit{(quasi)-periodic} signals are being \bedit{modeled} and removed. To do this, we compute the Lomb-Scargle Periodogram \citep{1976Ap&SS..39..447L, 1982ApJ...263..835S} of the RVs before and after applying the \calm\ activity correction. We used the periodogram functions in \texttt{astropy.timeseries} \citep{2012cidu.conf...47V, 2015ApJ...812...18V} to implement a generalized Lomb-Scargle Periodogram, where the periodograms are normalized according to the formalism in \citet{2009A&A...496..577Z}. In the periodograms corresponding to the old DRS (Figure \ref{fig:old_drs_correctedperiodogram}) and the new DRS (Figure \ref{fig:new_drs_correctedperiodogram}), we indicate the expected planet period (\litperiod\ days) from transits in yellow.  Based on the v$\sin{i} = 3.8$ km s$^{-1}$ derived from the \redit{SPC analysis of HARPS-N spectra}, we expect the rotation period of the star to be $\sim 15$ days and half the rotation period to be at $\sim 7.5$ days as indicated by $P_{\rm rot}$ and $P_{\rm rot}/2$ in grey in Figures \ref{fig:old_drs_correctedperiodogram}, \ref{fig:new_drs_correctedperiodogram}.

For the old DRS (3.7), we plot Lomb-Scargle Periodograms of the RVs before and after applying the activity correction models with increasing number of CCF \redit{indexes} in Figure \ref{fig:old_drs_correctedperiodogram}. As we increase the number of CCF \redit{indexes} (N) included in the \calm\ model from N=5 (Figure \ref{fig:old_drs_correctedperiodogram}b), to N = 15 (Figure \ref{fig:old_drs_correctedperiodogram}c), and finally to N = 25 (Figure \ref{fig:old_drs_correctedperiodogram}\redit{d}), we see that the peaks corresponding to $P_{\rm rot}$, $P_{\rm rot}/2$, and long-term stellar activity decrease in magnitude. A strong signal emerges instead at the the expected planet period (\litperiod\ days). Thus, a larger number of CCF \redit{indexes} (25) clearly more effectively separates stellar activity signals from \beditr{D}oppler reflex motion and reveals planetary signals. However, this does not imply \redit{that} we should infinitely increase the number of CCF \redit{indexes}. As discussed in Section \ref{no_more_overfitting}, increasing the number of CCF \redit{indexes} can result in too many free parameters compared to the number of observations and overfitting on noise in the data. We explored using 5-40 CCF \redit{indexes} and find that $\sim$ 25 CCF \redit{indexes} strikes the balance between preventing overfitting and still modeling the stellar activity signals effectively.

For the new DRS (2.3.5), we plot periodograms of the RVs before and after applying the activity correction model with 5 CCF \redit{indexes} in Figure \ref{fig:new_drs_correctedperiodogram}. We do not include several activity models with increasing number of CCF \redit{indexes} since we did not find that increasing the number of CCF \redit{indexes} made any notabl\redit{e} difference in the periodograms for the new DRS. Although we do see the magnitude of the peaks corresponding to $P_{\rm rot}$, $P_{\rm rot}/2$, and long-term stellar activity decrease marginally in magnitude and a slightly stronger peak corresponding to the planet period (\litperiod\ days) emerge in \ref{fig:new_drs_correctedperiodogram}b, this effect is much less pronounced compared to the old DRS. We expect that this is due to a difference in the use of linelists in the new and old DRS to compute the CCFs and RVs as discussed in Section \ref{mcmc_results}.


\subsection{Assessing the possibility of a second planetary companion}\label{2nd_companion}
In our activity corrected periodograms (bottom panels of Figures \ref{fig:old_drs_correctedperiodogram}, \ref{fig:new_drs_correctedperiodogram}), we noticed a peak arising at $\sim$ 22 days that could be hinting at a secondary companion. This period does not correspond to sidereal day, sidereal year, or lunation period aliases of \thisplanet\ or the star's rotation period and aliases. \bedit{Thus, w}e performed a 2 planet fit with our CCF \bedit{linear} activity model where \thisplanet\ is modeled as a Keplerian and the possible companion is modeled as a \bedit{cosine} function to keep free parameters to a minimum. Although our DE-MCMC model converged (Gelman Rubin values of < 1.009) and the fit looks relatively convincing as seen in Figure \ref{fig:two_planets_phasefold}, we do not find a significant detection for this potential second companion. From \bedit{the corner plot showing the distribution of possible periods (P2) in} Figure \ref{fig:two_planets}, we see that multiple periods between 21.6 and 22.8 could fit the observed data\bedit{, which may help explain the lack of significant detection}. Additional RV measurements are necessary to constrain the period and confirm or rule out this companion's existence.

\subsubsection{Comparing \calm\ to traditional activity indicators}

\begin{figure}
    \includegraphics[width=0.5\textwidth]{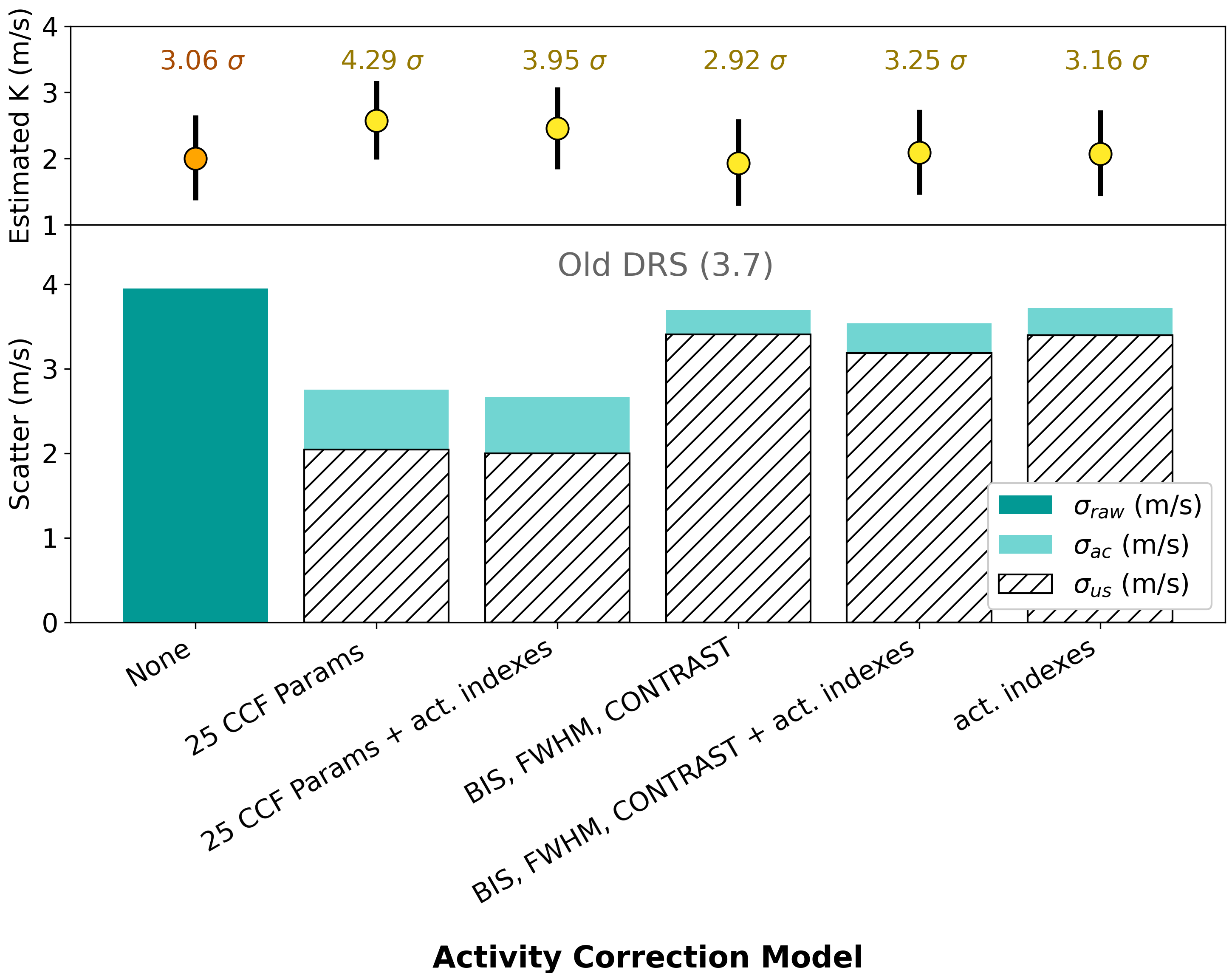}
    \caption{Comparison between using CCFs, CCF metrics (FWHM, bisector, contrast), and activity indexes (s-index, CaII, NaD, H$_{\alpha}$) as activity correction models for the Old DRS (3.7). On the bottom panel, the corrected scatter metrics ($\sigma_{\rm ac}$, $\sigma_{\rm us}$) for each of the activity models are plotted. The activity corrected standard deviation $\sigma_{\rm ac}$ measures the scatter in the activity corrected RV measurements. The unexplained scatter ($\sigma_{\rm us}$) measures the scatter remaining after removing all known contributions (stellar activity, planet signal, estimated instrumental noise, photon noise) to the RV signal. These scatter metrics are defined in more detail in  Section \ref{scatter_metrics}. In the top panel, we \bedit{include} the estimated semi-amplitude of \thisplanet\ for each activity correction model and the corresponding error bars and detection significance \bedit{to check how well the planet signal is preserved for each activity mitigation method}.}
    \label{fig:method_comparison}
\end{figure}

For the old DRS, which shows \redit{a} large\redit{r} scatter, we compared our CCF activity indicators with traditional activity indicators such as s-index, H$\alpha$, NaD, and other CCF metrics such as bisector (BIS), full-width-half-maximum (FWHM), and contrast. In Figure \ref{fig:method_comparison}, we evaluate these methods and compare them with the CCF method by both examining how well the planet signal is preserved and what the final remaining scatter in the RV measurements is. We use the scatter metrics defined in detail in Section \ref{scatter_metrics} for this comparison. We find that using any combination of activity indexes and/or BIS, FWHM results in a higher final unexplained scatter and a less significant detection of \thisplanet's mass. In particular, using only the BIS, FWHM, and contrast results in the highest final unexplained scatter of 3.18 \ms\ with the lowest significance of detection of 2.92$\sigma$. The other activity indicator model\redit{s} perform marginally better \redit{ and yield detections of 3.16$\sigma$ to 3.25$\sigma$}. \redit{However, when we include CCFs, we see a significant reduction in unexplained scatter and increased detection significance. Both the 25 CCF \calm\ model and the 25 CCF model with activity indexes (s-index, H$\alpha$, NaD)  yield a $\sigma_{\rm us} = 2.0$ \ms, which is the lowest $\sigma_{\rm us}$ we find. The 25 CCF \calm\ model has the higher detection significance with 4.29$\sigma$ compared to 3.95$\sigma$ for the CCF with activity indexes model. Thus, overall } the 25 CCF \calm\ model leads to the lowest $\sigma_{\rm us}$ and the highest detection significance among these methods.

\begin{table*}
\small
\begin{tabular}{lcc}\label{exofast_table}
Parameter & Units & Values
\\\hline
\multicolumn{3}{l}{Observed Stellar Parameters}\\
\hline\\
$\log{g}$\dotfill & Spectroscopic surface gravity (cgs)\dotfill & $4.10 \pm 0.05$ \\
$T_{\rm eff}$\dotfill &Effective Temperature (K)\dotfill & $6083 \pm 90 $ \\
$[{\rm Fe/H}]$\dotfill &Metallicity (dex)\dotfill & $-0.34 \pm 0.09 $\\
$v\sin{i}$\dotfill &Projected rotational velocity (\kms)\dotfill &$3.8 \pm 0.5$\\
\\\hline
\multicolumn{3}{l}{Derived Stellar Parameters}\\
\hline\\
$M_{\star}$\dotfill &Mass ($M_{\odot}$)\dotfill &$1.109^{+0.11}_{-0.091}$\\
$R_{\star}$\dotfill &Radius ($R_{\odot}$)\dotfill &$1.554^{+0.051}_{-0.050}$\\
$L_{\star}$\dotfill &Luminosity ($L_{\odot}$)\dotfill &$3.08^{+0.16}_{-0.17}$\\
$\rho_{\star}$\dotfill &Density (cgs)\dotfill &$0.419^{+0.065}_{-0.059}$\\
\textcolor{new_color}{$Age$}\dotfill &\redit{Age (Gyr)}$^{a}$\dotfill &\textcolor{new_color}{$5.6^{+2.2}_{-2.0}$}\\ 
$\log{g}$\dotfill &Model-derived surface gravity (cgs)\dotfill &$4.102^{+0.052}_{-0.055}$\\
$u_{1}$\dotfill &Kepler-band linear limb-darkening coeff \dotfill &$0.342\pm0.050$\\
$u_{1}$\dotfill &\TESS-band linear limb-darkening coeff \dotfill &$0.242\pm0.030$\\
$u_{2}$\dotfill &Kepler-band quadratic limb-darkening coeff \dotfill &$0.311^{+0.049}_{-0.050}$\\
$u_{2}$\dotfill &\TESS-band quadratic limb-darkening coeff \dotfill &$0.298\pm0.029$\\
\\\hline
\multicolumn{2}{l}{Planetary Parameters:}&b \\
\hline\\
$P$\dotfill &Period (days)\dotfill &$9.978543^{+0.000023}_{-0.000020}$\\
$M_P$\dotfill &Mass (\me)\dotfill & \erikamass\\
$R_P$\dotfill &Radius  (\re) $^{b}$\dotfill &\erikaradius\\
$T_C$\dotfill &Time of conjunction (\bjdtdb)$^{c}$\dotfill &$2456979.9329\pm0.0016$\\
$a$\dotfill &Semi-major axis (AU)\dotfill &$0.0939^{+0.0029}_{-0.0027}$\\
$i$\dotfill &Inclination (Degrees)\dotfill &$86.61^{+1.5}_{-0.53}$\\
$e$\dotfill &Eccentricity \dotfill &$0.30^{+0.16}_{-0.19}$\\
$\omega_{\star}$\dotfill &Argument of periastron (degrees) \dotfill &$99^{+44}_{-37}$\\
$e\cos{\omega_*}$\dotfill & \dotfill &$-0.03\pm0.14$\\
$e\sin{\omega_*}$\dotfill &\dotfill &$0.26^{+0.17}_{-0.23}$\\
$M_{P}$ $\sin i$\dotfill &Minimum mass (\me) \dotfill &$6.962^{+1.653}_{-1.621}$\\ 
$M_P/M_*$\dotfill&Mass ratio \dotfill &$0.0000173\pm0.0000044$\\
K\dotfill &RV semi-amplitude (\ms) \dotfill &\erikaK\\
$T_{eq}$\dotfill &Equilibrium temperature (K)$^{d}$\dotfill &$1203^{+25}_{-27}$\\
$R_P/R_\star$\dotfill &Radius of planet in stellar radii \dotfill &$0.01444^{+0.00077}_{-0.00057}$\\
$a/R_\star$\dotfill &Semi-major axis in stellar radii \dotfill &$13.01\pm0.64$\\
$d/R_\star$\dotfill &Planet/star separation at mid transit \dotfill &$9.5^{+3.0}_{-2.3}$\\
$\delta$\dotfill &Transit depth in Kepler $\left (R_p/R_\star \right )^2$\dotfill &$0.0002299^{+0.0000086}_{-0.0000085}$\\
$\delta$\dotfill &Transit depth in \TESS\ $\left (R_p/R_\star \right )^2$ \beditr{$^e$}\dotfill &$0.0002235^{+0.0000097}_{-0.0000096}$\\
$T_{14}$\dotfill &Total transit duration (days)\dotfill &$0.1507^{+0.0037}_{-0.0039}$\\
$b$\dotfill &Transit Impact parameter \dotfill &$0.60^{+0.18}_{-0.37}$\\
$\fave$\dotfill &Incident Flux (\fluxcgs)\dotfill &$0.425^{+0.054}_{-0.045}$\\
\end{tabular}
\caption{Photometric Properties and Median Values and 68\% Confidence Interval for Global EXOFASTv2 Models for the \thisstar\ system. See Table 3 in \citet{Eastman2019} for a detailed description of all parameters.}
a) \redit{We note that age measurements derived from isochrones have been found to have large systematic uncertainties \citep{Torres2010}.} \\b) We note that this is in  equatorial earth radii. \\ c) Time of conjunction is commonly reported as the "transit time" \\
d) Assumes no albedo and perfect redistribution. \\
e) The difference in transit depths for Kepler and \TESS\ is \beditrr{is due to limb darkening effects.}
\end{table*}

\subsection{Final RV model}

For our final RV model, we use the new DRS (2.3.5) RVs and the 5 CCF activity indicator \calm\ activity correction model since this yields the lowest final unexplained scatter ($\sigma_{\rm us}$). The CCF activity indicators are then used in the global transit, radial velocity, stellar parameter fit to decorrelate the RVs as further described in Section \ref{transit_analysis}. We do not include the possible additional planet that we investigated as described in Section \ref{2nd_companion} in our final RV model.

\section{\thisstar\ System Parameters}\label{system_parameters}
In addition to performing the RV activity mitigation analysis, we also measure the spectroscopic properties and use the CALM activity indicators to perform a global transit and stellar parameter analysis to derive the refined \thisstar\ system parameters as described in this section.

\subsection{Spectroscopic Parameters}\label{spectroscopic_analysis}
Using all 74 of the HARPS-N spectra, we measure the spectroscopic properties of \thisstar\ using three independent methods: ARES+MOOG\footnote{ARESv2: http://www.astro.up.pt/~sousasag/ares/}$^{,}$\footnote{MOOG-
2017: http://www.as.utexas.edu/~chris/moog.html} \citep{2014dapb.book..297S}, CCFpams\footnote{https://github.com/LucaMalavolta/CCFpams} \citep{2017MNRAS.469.3965M}, and Stellar Parameter Classification (SPC) \citep{2014Natur.509..593B}. 

First, we measured the spectroscopic parameters using ARES+MOOG\redit{, which is a curve-of-growth method that relies on neutral and ionized iron lines and is further} described by \citet{MortierrefId0}. We co-added all the spectra, used ARES2 \citep{2014dapb.book..297S,2015A&A...577A..67S} to measured equivalent widths of iron lines, and used MOOG \citep{1973ApJ...184..839S} to determine atmospheric parameters. We accounted for systematic effects in ARES+MOOG by using the surface gravity corrected method from \citet{2014A&A...572A..95M}. Our systematic errors were added to our precision errors in quadrature. These precision errors are intrinsic to the method. The systematic errors are 60 K for the the effective temperature, 0.1 dex for the surface gravity, and 0.04 dex for the metallicity \citep{2011A&A...533A.141S}. This analysis yielded an effective temperature $T_{\rm eff,MOOG} = 6115 \pm 69 $ K, surface gravity $\log g_{\rm cgs, MOOG} = 4.12 \pm 0.13$, and an iron abundance [Fe/H]$_{\rm MOOG} = -0.32 \pm 0.04$.

The CCFpams method \citep{2017MNRAS.469.3965M}, which relies on an empirical calibration that uses the width of CCFs to determine the effective temperature, surface gravity, and iron abundance. \redit{Using the same systematic errors for effective temperature, surface gravity, and metallicity as used for ARES+MOOG from \citet{2011A&A...533A.141S}, t}his method computed an  effective temperature $T_{\rm eff,CCFpams} =6092  \pm  \textcolor{new_color}{68}$ K, surface gravity $\log g_{\rm cgs, CCFpams} = 4.09 \pm \textcolor{new_color}{0.12} $, and an iron abundance [Fe/H]$_{\rm CCFpams} =-0.34 \pm  \textcolor{new_color}{0.05}$.

We used the Stellar Parameter Classification (SPC) method \citep{2014Natur.509..593B}, which determines stellar atmospheric parameters by cross-correlating a stellar spectrum with synthetic spectra from \citet{1992IAUS..149..225K} model atmospheres and then interpolating correlation peaks to find the effective temperature, surface gravity, and metallicity. We ran SPC on all our HARPS-N spectra. Using a prior on gravity of log $g_{\rm cgs, SPC} = 4.12$, we measured a temperature of $T_{\rm eff, SPC}=6047 \pm 49$K, a surface gravity of log $g_{\rm cgs, SPC} = 4.12 \pm 0.10$, an iron abundance $[Fe/H]_{\rm SPC} = -0.37 \pm 0.08$\redit{, and the projected rotational velocity v$\sin{i}_{\rm SPC} = 3.8 \pm 0.5$km s$^{-1}$}. 

\redit{Lastly, we computed the weighted average across all three methods. The weighted average yields an effective temperature $T_{\rm eff, avg}=6083 \pm 90$K, a surface gravity of log $g_{\rm cgs, avg} = 4.10 \pm 0.05$, and an iron abundance $[Fe/H]_{\rm avg} = -0.34 \pm 0.09$. We adopt these values and include $T_{\rm eff, avg}$ and $[Fe/H]_{\rm avg}$ as Gaussian priors in our global transit and stellar parameter analysis.}


\subsection{Global Transit/Stellar Parameter Analysis}\label{transit_analysis}
 To refine the ephemerides and system parameters of \thisplanet, we used the fitting software EXOFASTv2 \citep{ 2013PASP..125...83E, 2017ascl.soft10003E, 2019arXiv190709480E}, which allows us to combine transit observations from several missions with our spectroscopic parameters and archival photometry. In particular, we simultaneously fit the \TESS, \textit{K2} photometry, \bedit{RVs}, and stellar parameters (see Section \ref{data}). \bedit{In addition to including the RVs, we include the CCF-based activity indicators derived from the \calm\ model (described in Section \ref{ccf_inputs}) and use these to detrend the RVs in the EXOFASTv2 fit}. Since this detrending is identical to performing a linear regression, incorporating the CCF activity indicators in EXOFASTv2 is identical to using our \calm\ model for the RVs. In this fit, we constrain the possible mass, radius, and age of the host star according to the MESA Isochrones and Stellar Tracks (MIST) and stellar evolution models \citep{2011ApJS..192....3P, 2013ApJS..208....4P, 2015ApJS..220...15P, 2016ApJS..222....8D, 2016ApJ...823..102C}.

 We required a Gelman-Rubin statistic \citep{1992Gelman} of less than 1.01 and at least 1000 independent draws in each parameter to ensure strict convergence criteria. \redit{We imposed Gaussian priors on $[Fe/H]$ and $T_{\rm eff}$ using the weighted averages derived from our spectroscopic analysis described in Section \ref{spectroscopic_analysis}. }The results of our fit are shown in Figure \ref{fig:transits} and Table \ref{exofast_table}. In the EXOFASTv2 fit, we find $M_p = $ \erikamass \me\ and thus we find close agreement between the \calm\ estimate of mass (\calmmass \me) and the mass measured by \citet{Bonomo2023} (\bonomomass \me).

\subsection{Stellar variability signatures in the photometry}

\begin{figure}
    \includegraphics[width=1.05\columnwidth]{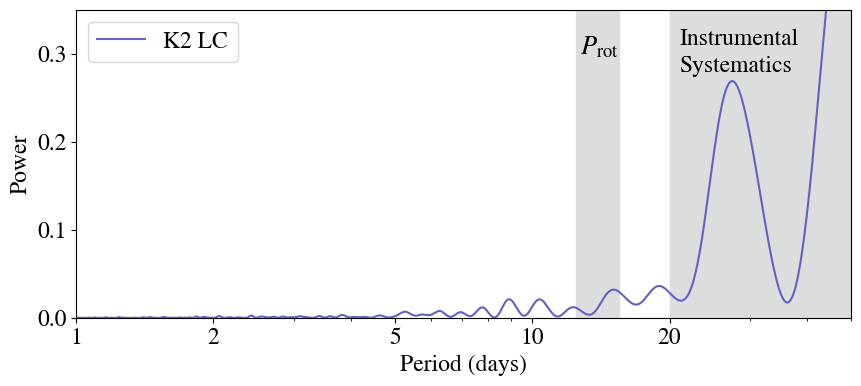}
    \caption{\rredit{Periodogram of K2 photometry. In grey, the expected stellar rotation period ($P_{\rm rot}$) of $\sim$ 15 days is indicated and the instrumental systematics that dominate signals with P $ > 20$ days are indicated. The instrumental systematics dominate the Fourier spectrum of the K2 light curve. }}
    \label{fig:k2_lc_periodogram}
\end{figure}

\rredit{We examined both the K2 and TESS light curves to look for stellar rotation signals. We did not find significant rotation signals in the TESS data. The K2 light curves have higher photometric precision and probe a bluer wavelength region compared to TESS and thus should be able to probe more stellar variability. For both the TESS and K2 data, We used the unflattened light curve that is not corrected for systematics since the systematic corrections often also remove stellar variability signals. We observe less than 0.1\% variability in the K2 light cuvres. In Figure \ref{fig:k2_lc_periodogram}, we examine the K2 light curve in Fourier space and find no significant signals at the expected rotation period ($P_{\rm rot} \sim 15$ days). We find that the periodogram is dominated by K2 instrumental systematics. Significant signal loss is expected for K2 due to long-term systematic noise (e.g. differential velocity aberration) for periods greater than 15 days \citep{VanCleve2016} and make it challenging to detect oscillations and rotation signals.}


\begin{figure*}
    \includegraphics[width=1\textwidth]{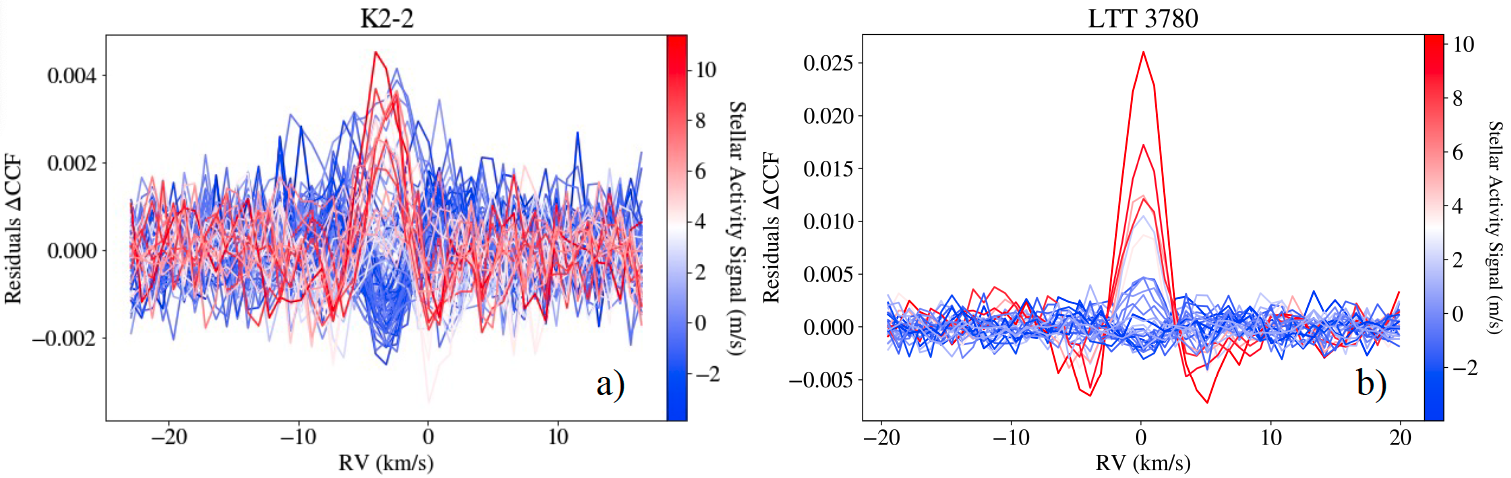}
    \caption{\rredit{Residual CCFs ($\Delta$CCFs) from the new DRS (2.3.5) for two example stars, K2-2 (Thygeson et al. in prep) and LTT 3780 \citep{Cloutier2020}. Although the new DRS (2.3.5) did not show a clear pattern in shape changes for \thisstar, we can see that other targets do clearly exhibit shape changes corresponding to stellar variability and could potentially benefit from our \calm\ method.}}
    \label{fig:other_stars}
\end{figure*}

\section{Discussion}\label{discussionconclusions}

\begin{figure*}
    \includegraphics[width=0.7\textwidth]{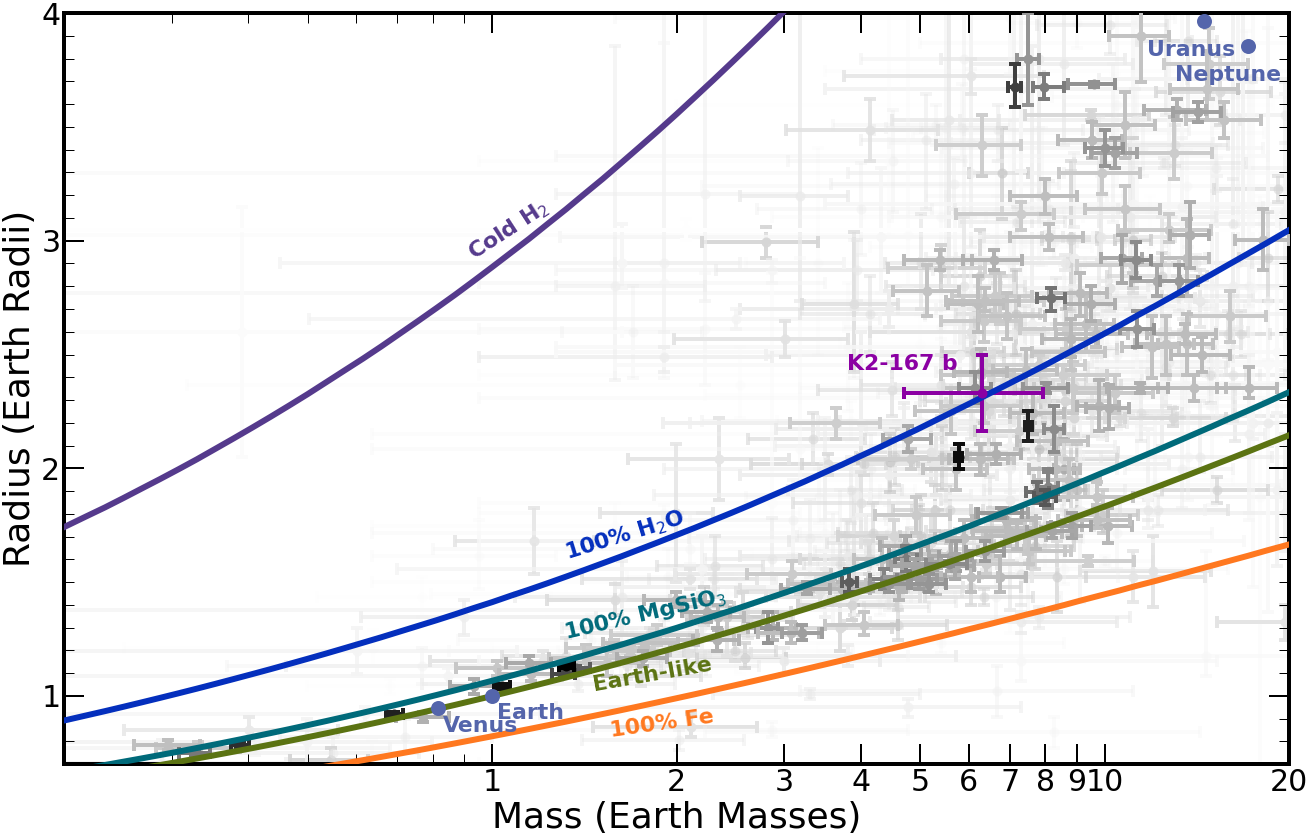}
    \caption{The mass/radius diagram for small exoplanets. Planet masses and radii come from the NASA Exoplanet Archive\citep{Akeson_2013}, accessed May 7, 2023.  The darkness of each of the symbols is directly proportional to the precision of the mass and radii measurements. The colored lines represent theoretical mass/radius relations for solid planets of various compositions from \citep{Zeng_2016} and for cold hydrogen planets from \citep{2007ApJ...669.1279S}. The planets in our solar system are plotted in blue and \thisplanet\ is plotted in purple with their corresponding uncertainties.  
    }
    \label{fig:mrdiagram}
\end{figure*}

In this work, we have measured the mass of \thisplanet\ as \calmmass\ \me\ by taking advantage of the reduction in scatter resulting from updates to the HARPS-N Data Reduction System (2.3.5) and our new stellar activity mitigation method (\calm). This measurement agrees within error with the recently published mass measurement of \citet{Bonomo2023} (\bonomomass \me), which was part of a larger statistical sample of exoplanet systems. Together with our radius measurement (\erikaradius \re), we can use our precise mass measurement to place constraints on \thisplanet's composition. We compared our measured mass and radius to mass-radius relations for planets of different compositions from  \citet{Zeng_2016}. In Figure \ref{fig:mrdiagram}, we demonstrate this newly determined mass in a mass/radius diagram. The mass and composition of \thisplanet\ is not consistent with an Earth-like composition (32.5\% Iron core, 67.5\% silicate mantle). Rather, \thisplanet\ is less dense than earth-like planets and falls closer to the pure ice (H$_2$O) mass-radius relation from \citet{Zeng_2016}.

This leaves several possibilities for the exact composition of \thisplanet. This planet could be a mixture of rock, iron, and hydrogen/helium (H/He). One possibility would be that \thisplanet\ has an earth-like iron core and mantle that is surrounded by a volatile envelope, which is \beditr{believed to be a} common composition for super-Earths and sub-Neptunes \citep{2014ApJ...783L...6W, rogers2015, 2015ApJ...800..135D}. Another possibility is that \thisplanet\ is a planet with a much smaller iron core than earth and mostly or entirely consists of water or methane or some other heavy volatiles.

Despite the fact that this planet was originally detected over 8 years ago, only recently has it been possible to make such inferences about its composition. Like many other low-mass planets, K2-167 b was first detected using the transit method and was validated several years ago but stellar activity prevented astronomers from measuring its mass until recently. To ensure the robustness of \thisplanet's mass measurement, \beditrr{we performed an extensive RV analysis where we compared our \calm\ framework to traditional activity mitigation methods (H-$\alpha$, s-index, FWHM, BIS, etc). Instead of using these indicators, our method exploits the activity-induced line shape changes in the spectra without requiring timing information like Gaussian Process regression. We trace these shape changes using CCFs and find that this method either outperforms or produces comparable results to traditional methods. }


The fact that different data reduction software versions\redit{, which use different linelists,} yield radial velocities with significantly distinct stellar activity characteristics suggests interesting future directions in terms of mitigating stellar activity.  \rredit{The effectiveness of spectra-based activity mitigation methods has already been shown to be highly dependent on linelists and McWilliam et al. in prep recently found that using chromatic CCFs, which are CCFs that use only part of the spectral orders, add complementary information for stellar activity mitigation for the Sun. Thus, in the future, we plan to generate our our line profiles generated using primarily activity sensitive lines such that we can apply \calm\ to these profiles rather than to CCFs from the HARPS-N DRS, which may greatly increase the effectiveness of spectra-based activity mitigation techniques like \calm.} 

\subsection{\rredit{Prospects for using \calm\ on other stars with the new DRS}}

\rredit{In the future, we plan to apply our \calm\ method to variety of others stars of different stellar types and different ages. Although the new DRS (2.3.5) RVs for \thisstar\ did not require our method to derive a mass a measurement, this is not the case for other targets. From preliminary analyses, we have found that  \calm\ can be effective at predicting and mitigating stellar variability for several stars with data from the new DRS (2.3.5). In Figure \ref{fig:other_stars}, we show two examples of the residual CCF observations for K2-2 (Thygeson at el. submitted), a K0-type star, and LTT 3790 \citep{Cloutier2020}, an M4-type star. Both these targets have planets confirmed with transits and we find that \calm\ can reduce the RV scatter while preserving and increasing the signal strength of the planetary reflex motion.}

\rredit{In addition to using \calm\ to model magnetic activity, we are hoping to develop methods to model granulation, which is especially relevant for stars with high surface gravity. The granulation-induced RV variation is anti-correlated with flicker, which are low-level photometric variations induced by granulation  \citep{Bastien2013}. Recently, \citet{Lakeland2023} found that granulation and supergranulation are some of the main remaining limiting factors in reaching tens of \cms\ level for solar observations. Thus, it is critical that we find ways to model and predict granulation and supergranulation, especially for solar-like stars.}

\subsection{\rredit{Applying \calm\ to young planetary systems}}
Improved activity mitigation methods are especially interesting for helping to constrain the mechanisms involved in creating the radius valley. One of the best ways to distinguish the most commonly proposed mechanisms, photoevaporation and core-powered mass loss, is by investigating the timescale on which mass loss takes place. Since photoevaporation is expected to proceed on timescales shorter than 100 million years, measuring the compositions of planets younger than this age is of strong interest. Unfortunately, stars this young typically have very high levels of stellar activity, making their precise characterization difficult \redit{and making them prone to overfitting using current state-of-the art techniques like Gaussian Processes \citep{Blunt2023}}. Applying \calm\ or similar techniques to these young stars could open the door to measuring more masses and densities for transiting planets orbiting bright young stars \redit{(i.e. EPIC 247589423 \citep{mann2018}, HD 283869 \citep{vanderburg2018}, V1298 Tau \citep{david2019}, DS Tuc A \citep{newton2019, Benatti2019})} \redit{and provide insight into the timescales of mass loss for these planets.}

\subsection{Future further characterization of \thisplanet}
\redit{In the future, additional RV observations of \thisplanet\ could allow us to use a larger portion of \rredit{the average line spectrum} and potentially improve our modeling of the stellar variability. If timed during transit, additional RV measurements could allow us to measure the spin-orbit angle using the Rossiter-McLaughlin (R-M) effect \citep{Rossiter1924, McLaughlin1924}. We compute the expected RM radial velocity deviation $\Delta RV_{\rm RM}$ using the approximation from \citet{Seager2010} that 
\begin{equation}
    \Delta RV_{\rm RM} = (\frac{Rp}{R_{\star}})^2 \sqrt{1-b^2} v sin i
\end{equation}
For \thisplanet, the expected RM radial velocity deviation is $\Delta RV_{\rm RM} \sim 20.1 $ \ms and should be measure-able with precision and high-precision spectographs. R-M measurements could provide insight into the angular momentum history and thereby dynamical history of the \thisstar\ system.} 

\redit{For future atmospheric target selection studies, we calculated the Transmission Spectroscopy Metric \citep[TSM;][]{Kempton2018} and the Emission Spectroscopy Metric \citep[ESM;][]{Kempton2018} for \thisplanet. We find that $TSM = 6.06$ and $ESM =  4.12$, which means that \thisplanet\ is not an ideal target for transmission and a moderately favorable target for emission.}

\section{\rredit{Conclusion}}\label{Conclusion}
\rredit{In this paper, we have developed a new spectral-based activity mitigation method called \calm\ and used it to characterize the \thisstar\ system, which has a planet at the edge of the radius valley. In-depth characterization of systems like this can be especially critical in probing the formation physics of the radius valley. In this paper, we have specifically obtained the following results:}

\rredit{(1) We performed a detailed RV analysis, where we compared \calm\ to various traditional activity indicators, evaluated various HARPS-N DRS versions, and tested a range of number of CCF activity indicators within the \calm\ framework. Across all of these analysis choices, we derive a consistent mass within uncertainty and find that the period agrees with the transit period of P = \litperiod\ days. Our final \calm\ model derives a mass of $M_p = $ \calmmass \me.}

\rredit{(2) We find that the effectiveness of \calm\ at reducing the RV scatter is highly dependent on the version of the HARPS-N DRS used, which suggests that this is highly dependent on the choice of linelists. In the future, custom linelists to compute CCFs may greatly increase the the effectiveness of spectra based techniques like \calm.}

\rredit{(3) We investigated our RV results in Fourier space using periodograms to understand which periodic signals were responsible for the observed decrease in RV scatter. Using \calm, we find that both stellar rotation signals and long-term magnetic activity signals decrease in magnitude while preserving the planet signal. This effect is most prominent in the old DRS (3.7). In the new DRS (2.3.5), we do not see the signal strength of activity signals signal decrease significantly, but the planet signal is preserved nonetheless. We expect that this difference is the result of the different linelists used in the different DRS versions.}

\rredit{(4) We performed a joint fit with the RVs and transit using EXOFASTv2. We found consistent mass estimates and refined the radius to $R_p = $ \erikaradius \re, which places \thisplanet\ at the upper edge of the radius valley. We refined the stellar parameters using three spectroscopic parameter estimation methods, and computed the weighted average across these methods, which yields an effective temperature $T_{\rm eff, avg}=6083 \pm 90$K, a surface gravity of log $g_{\rm cgs, avg} = 4.10 \pm 0.05$, and an iron abundance $[Fe/H]_{\rm avg} = -0.34 \pm 0.09$. }

\rredit{(5) We investigated the possibility of a secondary companion at a $\sim$22 day orbital period, but we found degeneracies in the possible periods and no significant detection. Further RV follow-up may help refine the period and confirm or rule out a possible companion.}

\rredit{(6) We introduced a new scatter metric called the unexplained scatter that accounts for both modelled stellar variability and modelled planetary signals to asses model performance. We introduce this metric since commonly used scatter metrics do not account for scatter originating from un-modelled planets and sometimes inadvertently incentivize removing planet signals rather than preserving them.}

\rredit{(7) We compared our \calm\ model to traditional activity indicators (s-index, H$\alpha$, NaD, bisector of the CCF, FWHM of the CCF, contrast of the CCF) and find that our models outperform any combination of these indicators for the old DRS. Our CCF-based \calm\ model both reduces the unexplained RV scatter and results in more significant mass detections.}

\rredit{(8) Combining our mass and radius measurements, we place constraints on \thisplanet's composition and find that it is less dense than would be expectected for an earth-like composition. This means that it could have either (i) an earth-like iron core and mantle with a volatile envelope, or (ii) a much smaller iron core and further consist of water, methane, or other heavy volatiles.}

\rredit{(9) Furthermore, we demonstrate that although \calm\ only yields a more significant mass detection for the old DRS (3.7) and not the new DRS (2.3.5), this is not the case for several other stars (e.g. K-type stars K2-2, M-type star LTT 3790) where we do reduce the RV scatter and increase the mass detection significance using the new DRS.}

In the future, continued development and broad deployment of stellar activity mitigation algorithms will have wide-ranging applications in exoplanet science. Methods like these could unlock the ability to measure masses  of planets orbiting active stars spanning the HR diagram and pave the way towards precise characterization of larger exoplanet populations in the super-Earth and sub-Neptune regime.

\section*{Acknowledgements}

We acknowledge helpful conversations and feedback from members of Dave Latham's Coffee Club. \beditr{We thank Dr. Sai Ravela for fruitful discussions on the best ways to choose the model input parameters and suggesting future directions for refining our process}. The HARPS-N project has been funded by the Prodex Program of the Swiss Space Office (SSO), the Harvard University Origins of Life Initiative (HUOLI), the Scottish Universities Physics Alliance (SUPA), the University of Geneva, the Smithsonian Astrophysical Observatory (SAO), and the Italian National Astrophysical Institute (INAF), the University of St Andrews, Queen's University Belfast, and the University of Edinburgh. 

ZLD would like to thank the generous support of the MIT Presidential Fellowship, \redit{the MIT Collamore-Rogers Fellowship} and to acknowledge that this material is based upon work supported by the National Science Foundation Graduate Research Fellowship under Grant No. 1745302.  ZLD and AV acknowledge support from the TESS Guest Investigator Program under NASA grant 80NSSC19K0388. ZLD and AV acknowledge support from the D.17 Extreme Precision Radial Velocity Foundation Science program under NASA grant 80NSSC22K0848.

\redit{AMS would like to acknowledge that this work was supported by FCT - Fundação para a Ciência e a Tecnologia through national funds by these grants: UIDB/04434/2020; UIDP/04434/2020". Funded/Co-funded by the European Union (ERC, FIERCE, 101052347). Views and opinions expressed are however those of the author(s) only and do not necessarily reflect those of the European Union or the European Research Council. Neither the European Union nor the granting authority can be held responsible for them. A.M.S acknowledges support from the Fundação para a Ciência e a Tecnologia (FCT) through the Fellowship 2020.05387.BD.}

\redit{R.D.H. is funded by the UK Science and Technology Facilities Council (STFC)'s Ernest Rutherford Fellowship (grant number ST/V004735/1).}

This research has made use of the NASA Exoplanet Archive, which is operated by the California Institute of Technology, under contract with the National Aeronautics and Space Administration under the Exoplanet Exploration Program.
\section*{Data Availability}

All RVs and activity indicators used in our analyses will be available via VizieR at CDS. \redit{The \TESS\ and \Ktwo\ photometry data are accesible via the Mikulski Archive
for Space Telescopes portal at \url{https://mast.stsci.edu/portal/
Mashup/Clients/Mast/Portal.html}.}



\bibliographystyle{mnras}
\bibliography{bib} 

\begin{thebibliography}{}
\makeatletter
\relax
\def\mn@urlcharsother{\let\do\@makeother \do\$\do\&\do\#\do\^\do\_\do\%\do\~}
\def\mn@doi{\begingroup\mn@urlcharsother \@ifnextchar [ {\mn@doi@}
  {\mn@doi@[]}}
\def\mn@doi@[#1]#2{\def\@tempa{#1}\ifx\@tempa\@empty \href
  {http://dx.doi.org/#2} {doi:#2}\else \href {http://dx.doi.org/#2} {#1}\fi
  \endgroup}
\def\mn@eprint#1#2{\mn@eprint@#1:#2::\@nil}
\def\mn@eprint@arXiv#1{\href {http://arxiv.org/abs/#1} {{\tt arXiv:#1}}}
\def\mn@eprint@dblp#1{\href {http://dblp.uni-trier.de/rec/bibtex/#1.xml}
  {dblp:#1}}
\def\mn@eprint@#1:#2:#3:#4\@nil{\def\@tempa {#1}\def\@tempb {#2}\def\@tempc
  {#3}\ifx \@tempc \@empty \let \@tempc \@tempb \let \@tempb \@tempa \fi \ifx
  \@tempb \@empty \def\@tempb {arXiv}\fi \@ifundefined
  {mn@eprint@\@tempb}{\@tempb:\@tempc}{\expandafter \expandafter \csname
  mn@eprint@\@tempb\endcsname \expandafter{\@tempc}}}

\bibitem[\protect\citeauthoryear{Akeson et~al.,}{Akeson
  et~al.}{2013}]{Akeson_2013}
Akeson R.~L.,  et~al., 2013, \mn@doi [Publications of the Astronomical Society
  of the Pacific] {10.1086/672273}, 125, 989

\bibitem[\protect\citeauthoryear{{Barrag{\'a}n} et~al.,}{{Barrag{\'a}n}
  et~al.}{2019}]{Barragan2019}
{Barrag{\'a}n} O.,  et~al., 2019, \mn@doi [\mnras] {10.1093/mnras/stz2569},
  \href {https://ui.adsabs.harvard.edu/abs/2019MNRAS.490..698B} {490, 698}

\bibitem[\protect\citeauthoryear{{Barrag{\'a}n} et~al.,}{{Barrag{\'a}n}
  et~al.}{2023}]{Barragan2023}
{Barrag{\'a}n} O.,  et~al., 2023, \mn@doi [\mnras] {10.1093/mnras/stad1139},
  \href {https://ui.adsabs.harvard.edu/abs/2023MNRAS.522.3458B} {522, 3458}

\bibitem[\protect\citeauthoryear{{Bastien}, {Stassun}, {Basri}  \&
  {Pepper}}{{Bastien} et~al.}{2013}]{Bastien2013}
{Bastien} F.~A.,  {Stassun} K.~G.,  {Basri} G.,   {Pepper} J.,  2013, \mn@doi
  [\nat] {10.1038/nature12419}, \href
  {https://ui.adsabs.harvard.edu/abs/2013Natur.500..427B} {500, 427}

\bibitem[\protect\citeauthoryear{{Batalha} et~al.,}{{Batalha}
  et~al.}{2013}]{batalha2013}
{Batalha} N.~M.,  et~al., 2013, \mn@doi [\apjs] {10.1088/0067-0049/204/2/24},
  \href {https://ui.adsabs.harvard.edu/abs/2013ApJS..204...24B} {204, 24}

\bibitem[\protect\citeauthoryear{{Benatti} et~al.,}{{Benatti}
  et~al.}{2019}]{Benatti2019}
{Benatti} S.,  et~al., 2019, \mn@doi [\aap] {10.1051/0004-6361/201935598},
  \href {https://ui.adsabs.harvard.edu/abs/2019A&A...630A..81B} {630, A81}

\bibitem[\protect\citeauthoryear{{Berger}, {Huber}, {Gaidos}  \& {van
  Saders}}{{Berger} et~al.}{2018}]{2018Berger}
{Berger} T.~A.,  {Huber} D.,  {Gaidos} E.,   {van Saders} J.~L.,  2018, \mn@doi
  [\apj] {10.3847/1538-4357/aada83}, \href
  {https://ui.adsabs.harvard.edu/abs/2018ApJ...866...99B} {866, 99}

\bibitem[\protect\citeauthoryear{{Blunt} et~al.,}{{Blunt}
  et~al.}{2023}]{Blunt2023}
{Blunt} S.,  et~al., 2023, \mn@doi [\aj] {10.3847/1538-3881/acde78}, \href
  {https://ui.adsabs.harvard.edu/abs/2023AJ....166...62B} {166, 62}

\bibitem[\protect\citeauthoryear{{Boisse} et~al.,}{{Boisse}
  et~al.}{2009}]{Boisse2009}
{Boisse} I.,  et~al., 2009, \mn@doi [\aap] {10.1051/0004-6361:200810648}, \href
  {https://ui.adsabs.harvard.edu/abs/2009A&A...495..959B} {495, 959}

\bibitem[\protect\citeauthoryear{{Bonfils} et~al.,}{{Bonfils}
  et~al.}{2007}]{2007A&A...474..293B}
{Bonfils} X.,  et~al., 2007, \mn@doi [\aap] {10.1051/0004-6361:20077068}, \href
  {https://ui.adsabs.harvard.edu/abs/2007A&A...474..293B} {474, 293}

\bibitem[\protect\citeauthoryear{{Bonomo} et~al.,}{{Bonomo}
  et~al.}{2023}]{Bonomo2023}
{Bonomo} A.~S.,  et~al., 2023, \mn@doi [arXiv e-prints]
  {10.48550/arXiv.2304.05773}, \href
  {https://ui.adsabs.harvard.edu/abs/2023arXiv230405773B} {p. arXiv:2304.05773}

\bibitem[\protect\citeauthoryear{{Borucki} et~al.,}{{Borucki}
  et~al.}{2010}]{Borucki2010}
{Borucki} W.~J.,  et~al., 2010, \mn@doi [Science] {10.1126/science.1185402},
  \href {https://ui.adsabs.harvard.edu/abs/2010Sci...327..977B} {327, 977}

\bibitem[\protect\citeauthoryear{{Buchhave} et~al.,}{{Buchhave}
  et~al.}{2014}]{2014Natur.509..593B}
{Buchhave} L.~A.,  et~al., 2014, \mn@doi [\nat] {10.1038/nature13254}, \href
  {https://ui.adsabs.harvard.edu/abs/2014Natur.509..593B} {509, 593}

\bibitem[\protect\citeauthoryear{{Choi}, {Dotter}, {Conroy}, {Cantiello},
  {Paxton}  \& {Johnson}}{{Choi} et~al.}{2016}]{2016ApJ...823..102C}
{Choi} J.,  {Dotter} A.,  {Conroy} C.,  {Cantiello} M.,  {Paxton} B.,
  {Johnson} B.~D.,  2016, \mn@doi [\apj] {10.3847/0004-637X/823/2/102}, \href
  {https://ui.adsabs.harvard.edu/abs/2016ApJ...823..102C} {823, 102}

\bibitem[\protect\citeauthoryear{{Cloutier} et~al.,}{{Cloutier}
  et~al.}{2020}]{Cloutier2020}
{Cloutier} R.,  et~al., 2020, \mn@doi [\aj] {10.3847/1538-3881/ab91c2}, \href
  {https://ui.adsabs.harvard.edu/abs/2020AJ....160....3C} {160, 3}

\bibitem[\protect\citeauthoryear{{Colwell}, {Timmaraju}  \& {Wise}}{{Colwell}
  et~al.}{2023}]{Colwell2023}
{Colwell} I.,  {Timmaraju} V.,   {Wise} A.,  2023, \mn@doi [arXiv e-prints]
  {10.48550/arXiv.2304.04807}, \href
  {https://ui.adsabs.harvard.edu/abs/2023arXiv230404807C} {p. arXiv:2304.04807}

\bibitem[\protect\citeauthoryear{{Cosentino} et~al.,}{{Cosentino}
  et~al.}{2012}]{2012SPIE.8446E..1VC}
{Cosentino} R.,  et~al., 2012, in \procspie. p. 84461V,
  \mn@doi{10.1117/12.925738}

\bibitem[\protect\citeauthoryear{{Dattilo}, {Batalha}  \& {Bryson}}{{Dattilo}
  et~al.}{2023}]{dattilo2023}
{Dattilo} A.,  {Batalha} N.~M.,   {Bryson} S.,  2023, \mn@doi [arXiv e-prints]
  {10.48550/arXiv.2308.00103}, \href
  {https://ui.adsabs.harvard.edu/abs/2023arXiv230800103D} {p. arXiv:2308.00103}

\bibitem[\protect\citeauthoryear{{David} et~al.,}{{David}
  et~al.}{2019}]{david2019}
{David} T.~J.,  et~al., 2019, \mn@doi [\aj] {10.3847/1538-3881/ab290f}, \href
  {https://ui.adsabs.harvard.edu/abs/2019AJ....158...79D} {158, 79}

\bibitem[\protect\citeauthoryear{{Delisle} et~al.,}{{Delisle}
  et~al.}{2018}]{2018A&A...614A.133D}
{Delisle} J.~B.,  et~al., 2018, \mn@doi [\aap] {10.1051/0004-6361/201732529},
  \href {https://ui.adsabs.harvard.edu/abs/2018A&A...614A.133D} {614, A133}

\bibitem[\protect\citeauthoryear{{Deming} et~al.,}{{Deming}
  et~al.}{2015}]{Deming2015}
{Deming} D.,  et~al., 2015, \mn@doi [\apj] {10.1088/0004-637X/805/2/132}, \href
  {https://ui.adsabs.harvard.edu/abs/2015ApJ...805..132D} {805, 132}

\bibitem[\protect\citeauthoryear{{Dotter}}{{Dotter}}{2016}]{2016ApJS..222....8D}
{Dotter} A.,  2016, \mn@doi [\apjs] {10.3847/0067-0049/222/1/8}, \href
  {https://ui.adsabs.harvard.edu/abs/2016ApJS..222....8D} {222, 8}

\bibitem[\protect\citeauthoryear{{Dressing} et~al.,}{{Dressing}
  et~al.}{2015}]{2015ApJ...800..135D}
{Dressing} C.~D.,  et~al., 2015, \mn@doi [\apj] {10.1088/0004-637X/800/2/135},
  \href {https://ui.adsabs.harvard.edu/abs/2015ApJ...800..135D} {800, 135}

\bibitem[\protect\citeauthoryear{{Duck}, {Harada}, {Harrell}, {Morris},
  {Williams}, {Crossfield}, {Werner}  \& {Deming}}{{Duck}
  et~al.}{2021}]{Duck2021}
{Duck} A.,  {Harada} C.~K.,  {Harrell} J.,  {Morris} R. R.~A.,  {Williams} E.,
  {Crossfield} I.,  {Werner} M.,   {Deming} D.,  2021, \mn@doi [\aj]
  {10.3847/1538-3881/ac0e2f}, \href
  {https://ui.adsabs.harvard.edu/abs/2021AJ....162..136D} {162, 136}

\bibitem[\protect\citeauthoryear{{Dumusque} et~al.,}{{Dumusque}
  et~al.}{2011}]{Dumusque2011}
{Dumusque} X.,  et~al., 2011, \mn@doi [\aap] {10.1051/0004-6361/201117148},
  \href {https://ui.adsabs.harvard.edu/abs/2011A&A...535A..55D} {535, A55}

\bibitem[\protect\citeauthoryear{{Dumusque} et~al.,}{{Dumusque}
  et~al.}{2021}]{Dumusque2021}
{Dumusque} X.,  et~al., 2021, \mn@doi [\aap] {10.1051/0004-6361/202039350},
  \href {https://ui.adsabs.harvard.edu/abs/2021A&A...648A.103D} {648, A103}

\bibitem[\protect\citeauthoryear{{Eastman}}{{Eastman}}{2017}]{2017ascl.soft10003E}
{Eastman} J.,  2017, {EXOFASTv2: Generalized publication-quality exoplanet
  modeling code} (\mn@eprint {ascl} {1710.003})

\bibitem[\protect\citeauthoryear{{Eastman}, {Gaudi}  \& {Agol}}{{Eastman}
  et~al.}{2013}]{2013PASP..125...83E}
{Eastman} J.,  {Gaudi} B.~S.,   {Agol} E.,  2013, \mn@doi [\pasp]
  {10.1086/669497}, \href
  {https://ui.adsabs.harvard.edu/abs/2013PASP..125...83E} {125, 83}

\bibitem[\protect\citeauthoryear{{Eastman} et~al.,}{{Eastman}
  et~al.}{2019a}]{Eastman2019}
{Eastman} J.~D.,  et~al., 2019a, \mn@doi [arXiv e-prints]
  {10.48550/arXiv.1907.09480}, \href
  {https://ui.adsabs.harvard.edu/abs/2019arXiv190709480E} {p. arXiv:1907.09480}

\bibitem[\protect\citeauthoryear{{Eastman} et~al.,}{{Eastman}
  et~al.}{2019b}]{2019arXiv190709480E}
{Eastman} J.~D.,  et~al., 2019b, arXiv e-prints, \href
  {https://ui.adsabs.harvard.edu/abs/2019arXiv190709480E} {p. arXiv:1907.09480}

\bibitem[\protect\citeauthoryear{{Fazio} et~al.,}{{Fazio}
  et~al.}{2004}]{Fazio2004}
{Fazio} G.~G.,  et~al., 2004, \mn@doi [\apjs] {10.1086/422843}, \href
  {https://ui.adsabs.harvard.edu/abs/2004ApJS..154...10F} {154, 10}

\bibitem[\protect\citeauthoryear{{Figueira}, {Santos}, {Pepe}, {Lovis}  \&
  {Nardetto}}{{Figueira} et~al.}{2013}]{Figueira2013}
{Figueira} P.,  {Santos} N.~C.,  {Pepe} F.,  {Lovis} C.,   {Nardetto} N.,
  2013, \mn@doi [\aap] {10.1051/0004-6361/201220779}, \href
  {https://ui.adsabs.harvard.edu/abs/2013A&A...557A..93F} {557, A93}

\bibitem[\protect\citeauthoryear{{Fulton} et~al.,}{{Fulton}
  et~al.}{2017}]{2017Fulton}
{Fulton} B.~J.,  et~al., 2017, \mn@doi [\aj] {10.3847/1538-3881/aa80eb}, \href
  {https://ui.adsabs.harvard.edu/abs/2017AJ....154..109F} {154, 109}

\bibitem[\protect\citeauthoryear{{Fulton}, {Petigura}, {Blunt}  \&
  {Sinukoff}}{{Fulton} et~al.}{2018}]{Fulton2018}
{Fulton} B.~J.,  {Petigura} E.~A.,  {Blunt} S.,   {Sinukoff} E.,  2018, \mn@doi
  [\pasp] {10.1088/1538-3873/aaaaa8}, \href
  {https://ui.adsabs.harvard.edu/abs/2018PASP..130d4504F} {130, 044504}

\bibitem[\protect\citeauthoryear{{Garhart} et~al.,}{{Garhart}
  et~al.}{2020}]{Garhart2020}
{Garhart} E.,  et~al., 2020, \mn@doi [\aj] {10.3847/1538-3881/ab6cff}, \href
  {https://ui.adsabs.harvard.edu/abs/2020AJ....159..137G} {159, 137}

\bibitem[\protect\citeauthoryear{{Gelman} \& {Rubin}}{{Gelman} \&
  {Rubin}}{1992}]{1992Gelman}
{Gelman} A.,  {Rubin} D.~B.,  1992, \mn@doi [Statistical Science]
  {10.1214/ss/1177011136}, \href
  {https://ui.adsabs.harvard.edu/abs/1992StaSc...7..457G} {7, 457}

\bibitem[\protect\citeauthoryear{{Gilbertson}, {Ford}, {Jones}  \&
  {Stenning}}{{Gilbertson} et~al.}{2020}]{Gilbertson2020}
{Gilbertson} C.,  {Ford} E.~B.,  {Jones} D.~E.,   {Stenning} D.~C.,  2020,
  \mn@doi [\apj] {10.3847/1538-4357/abc627}, \href
  {https://ui.adsabs.harvard.edu/abs/2020ApJ...905..155G} {905, 155}

\bibitem[\protect\citeauthoryear{{Ginzburg}, {Schlichting}  \&
  {Sari}}{{Ginzburg} et~al.}{2018}]{ginzburg}
{Ginzburg} S.,  {Schlichting} H.~E.,   {Sari} R.,  2018, \mn@doi [\mnras]
  {10.1093/mnras/sty290}, \href
  {https://ui.adsabs.harvard.edu/abs/2018MNRAS.476..759G} {476, 759}

\bibitem[\protect\citeauthoryear{{Gupta} \& {Schlichting}}{{Gupta} \&
  {Schlichting}}{2020}]{gupta}
{Gupta} A.,  {Schlichting} H.~E.,  2020, \mn@doi [\mnras]
  {10.1093/mnras/staa315}, \href
  {https://ui.adsabs.harvard.edu/abs/2020MNRAS.493..792G} {493, 792}

\bibitem[\protect\citeauthoryear{{Haywood} et~al.,}{{Haywood}
  et~al.}{2014}]{2014MNRAS.443.2517H}
{Haywood} R.~D.,  et~al., 2014, \mn@doi [\mnras] {10.1093/mnras/stu1320}, \href
  {https://ui.adsabs.harvard.edu/abs/2014MNRAS.443.2517H} {443, 2517}

\bibitem[\protect\citeauthoryear{{Haywood} et~al.,}{{Haywood}
  et~al.}{2020}]{2020arXiv200513386H}
{Haywood} R.~D.,  et~al., 2020, arXiv e-prints, \href
  {https://ui.adsabs.harvard.edu/abs/2020arXiv200513386H} {p. arXiv:2005.13386}

\bibitem[\protect\citeauthoryear{{Haywood} et~al.,}{{Haywood}
  et~al.}{2022}]{Haywood2022}
{Haywood} R.~D.,  et~al., 2022, \mn@doi [\apj] {10.3847/1538-4357/ac7c12},
  \href {https://ui.adsabs.harvard.edu/abs/2022ApJ...935....6H} {935, 6}

\bibitem[\protect\citeauthoryear{{H{\'e}brard} et~al.,}{{H{\'e}brard}
  et~al.}{2010}]{Hebrard2010}
{H{\'e}brard} G.,  et~al., 2010, \mn@doi [\aap] {10.1051/0004-6361/200913525},
  \href {https://ui.adsabs.harvard.edu/abs/2010A&A...512A..46H} {512, A46}

\bibitem[\protect\citeauthoryear{{Householder} \& {Weiss}}{{Householder} \&
  {Weiss}}{2022}]{Householder2022}
{Householder} A.,  {Weiss} L.,  2022, \mn@doi [arXiv e-prints]
  {10.48550/arXiv.2212.06966}, \href
  {https://ui.adsabs.harvard.edu/abs/2022arXiv221206966H} {p. arXiv:2212.06966}

\bibitem[\protect\citeauthoryear{{Ikwut-Ukwa} et~al.,}{{Ikwut-Ukwa}
  et~al.}{2020}]{2020AJ....160..209I}
{Ikwut-Ukwa} M.,  et~al., 2020, \mn@doi [\aj] {10.3847/1538-3881/aba964}, \href
  {https://ui.adsabs.harvard.edu/abs/2020AJ....160..209I} {160, 209}

\bibitem[\protect\citeauthoryear{{Jenkins}}{{Jenkins}}{2002}]{2002ApJ...575..493J}
{Jenkins} J.~M.,  2002, \mn@doi [\apj] {10.1086/341136}, \href
  {https://ui.adsabs.harvard.edu/abs/2002ApJ...575..493J} {575, 493}

\bibitem[\protect\citeauthoryear{{Jenkins} et~al.,}{{Jenkins}
  et~al.}{2016}]{2016SPIE.9913E..3EJ}
{Jenkins} J.~M.,  et~al., 2016, in {Chiozzi} G.,  {Guzman} J.~C.,  eds,
  Society of Photo-Optical Instrumentation Engineers (SPIE) Conference Series
  Vol. 9913, Software and Cyberinfrastructure for Astronomy IV. p. 99133E,
  \mn@doi{10.1117/12.2233418}

\bibitem[\protect\citeauthoryear{{Jin}, {Mordasini}, {Parmentier}, {van
  Boekel}, {Henning}  \& {Ji}}{{Jin} et~al.}{2014}]{2014Jin}
{Jin} S.,  {Mordasini} C.,  {Parmentier} V.,  {van Boekel} R.,  {Henning} T.,
  {Ji} J.,  2014, \mn@doi [\apj] {10.1088/0004-637X/795/1/65}, \href
  {https://ui.adsabs.harvard.edu/abs/2014ApJ...795...65J} {795, 65}

\bibitem[\protect\citeauthoryear{{Jones}, {Stenning}, {Ford}, {Wolpert},
  {Loredo}  \& {Dumusque}}{{Jones} et~al.}{2017}]{2017arXiv171101318J}
{Jones} D.~E.,  {Stenning} D.~C.,  {Ford} E.~B.,  {Wolpert} R.~L.,  {Loredo}
  T.~J.,   {Dumusque} X.,  2017, arXiv e-prints, \href
  {https://ui.adsabs.harvard.edu/abs/2017arXiv171101318J} {p. arXiv:1711.01318}

\bibitem[\protect\citeauthoryear{{Kempton} et~al.,}{{Kempton}
  et~al.}{2018}]{Kempton2018}
{Kempton} E. M.~R.,  et~al., 2018, \mn@doi [\pasp] {10.1088/1538-3873/aadf6f},
  \href {https://ui.adsabs.harvard.edu/abs/2018PASP..130k4401K} {130, 114401}

\bibitem[\protect\citeauthoryear{{Kurucz}}{{Kurucz}}{1992}]{1992IAUS..149..225K}
{Kurucz} R.~L.,  1992, in {Barbuy} B.,  {Renzini} A.,  eds, The Stellar
  Populations of Galaxies. 149.
p.~225

\bibitem[\protect\citeauthoryear{{Lakeland} et~al.,}{{Lakeland}
  et~al.}{2023}]{Lakeland2023}
{Lakeland} B.~S.,  et~al., 2023, \mn@doi [\mnras] {10.1093/mnras/stad3723},
  \href {https://ui.adsabs.harvard.edu/abs/2023MNRAS.tmp.3567L} {}

\bibitem[\protect\citeauthoryear{{Lee}, {Karalis}  \& {Thorngren}}{{Lee}
  et~al.}{2022}]{lee}
{Lee} E.~J.,  {Karalis} A.,   {Thorngren} D.~P.,  2022, \mn@doi [\apj]
  {10.3847/1538-4357/ac9c66}, \href
  {https://ui.adsabs.harvard.edu/abs/2022ApJ...941..186L} {941, 186}

\bibitem[\protect\citeauthoryear{{Lienhard}, {Mortier}, {Cegla}, {Cameron},
  {Klein}  \& {Watson}}{{Lienhard} et~al.}{2023}]{Lienhard2023}
{Lienhard} F.,  {Mortier} A.,  {Cegla} H.~M.,  {Cameron} A.~C.,  {Klein} B.,
  {Watson} C.~A.,  2023, \mn@doi [\mnras] {10.1093/mnras/stad1343}, \href
  {https://ui.adsabs.harvard.edu/abs/2023MNRAS.522.5862L} {522, 5862}

\bibitem[\protect\citeauthoryear{{Lightkurve Collaboration}
  et~al.,}{{Lightkurve Collaboration} et~al.}{2018}]{2018ascl.soft12013L}
{Lightkurve Collaboration} et~al., 2018, {Lightkurve: Kepler and TESS time
  series analysis in Python} (\mn@eprint {ascl} {1812.013})

\bibitem[\protect\citeauthoryear{{Ligi} et~al.,}{{Ligi}
  et~al.}{2018}]{2018AJ....156..182L}
{Ligi} R.,  et~al., 2018, \mn@doi [\aj] {10.3847/1538-3881/aadc69}, \href
  {https://ui.adsabs.harvard.edu/abs/2018AJ....156..182L} {156, 182}

\bibitem[\protect\citeauthoryear{{Lomb}}{{Lomb}}{1976}]{1976Ap&SS..39..447L}
{Lomb} N.~R.,  1976, \mn@doi [\apss] {10.1007/BF00648343}, \href
  {https://ui.adsabs.harvard.edu/abs/1976Ap&SS..39..447L} {39, 447}

\bibitem[\protect\citeauthoryear{{Lopez} \& {Fortney}}{{Lopez} \&
  {Fortney}}{2013}]{2013Lopez}
{Lopez} E.~D.,  {Fortney} J.~J.,  2013, \mn@doi [\apj]
  {10.1088/0004-637X/776/1/2}, \href
  {https://ui.adsabs.harvard.edu/abs/2013ApJ...776....2L} {776, 2}

\bibitem[\protect\citeauthoryear{{Lovis} et~al.,}{{Lovis}
  et~al.}{2011}]{Lovis2011}
{Lovis} C.,  et~al., 2011, \mn@doi [arXiv e-prints] {10.48550/arXiv.1107.5325},
  \href {https://ui.adsabs.harvard.edu/abs/2011arXiv1107.5325L} {p.
  arXiv:1107.5325}

\bibitem[\protect\citeauthoryear{{Malavolta}, {Lovis}, {Pepe}, {Sneden}  \&
  {Udry}}{{Malavolta} et~al.}{2017}]{2017MNRAS.469.3965M}
{Malavolta} L.,  {Lovis} C.,  {Pepe} F.,  {Sneden} C.,   {Udry} S.,  2017,
  \mn@doi [\mnras] {10.1093/mnras/stx1100}, \href
  {https://ui.adsabs.harvard.edu/abs/2017MNRAS.469.3965M} {469, 3965}

\bibitem[\protect\citeauthoryear{{Mann} et~al.,}{{Mann}
  et~al.}{2018}]{mann2018}
{Mann} A.~W.,  et~al., 2018, \mn@doi [\aj] {10.3847/1538-3881/aa9791}, \href
  {https://ui.adsabs.harvard.edu/abs/2018AJ....155....4M} {155, 4}

\bibitem[\protect\citeauthoryear{{Mayo} et~al.,}{{Mayo}
  et~al.}{2018}]{2018AJ....155..136M}
{Mayo} A.~W.,  et~al., 2018, \mn@doi [\aj] {10.3847/1538-3881/aaadff}, \href
  {https://ui.adsabs.harvard.edu/abs/2018AJ....155..136M} {155, 136}

\bibitem[\protect\citeauthoryear{{Mayor} et~al.,}{{Mayor}
  et~al.}{2003}]{mayor2003}
{Mayor} M.,  et~al., 2003, The Messenger, \href
  {https://ui.adsabs.harvard.edu/abs/2003Msngr.114...20M} {114, 20}

\bibitem[\protect\citeauthoryear{{Mayor} et~al.,}{{Mayor}
  et~al.}{2011}]{mayor2011}
{Mayor} M.,  et~al., 2011, \mn@doi [arXiv e-prints] {10.48550/arXiv.1109.2497},
  \href {https://ui.adsabs.harvard.edu/abs/2011arXiv1109.2497M} {p.
  arXiv:1109.2497}

\bibitem[\protect\citeauthoryear{{McLaughlin}}{{McLaughlin}}{1924}]{McLaughlin1924}
{McLaughlin} D.~B.,  1924, \mn@doi [\apj] {10.1086/142826}, \href
  {https://ui.adsabs.harvard.edu/abs/1924ApJ....60...22M} {60, 22}

\bibitem[\protect\citeauthoryear{{Meunier} \& {Lagrange}}{{Meunier} \&
  {Lagrange}}{2013}]{Meunier2013}
{Meunier} N.,  {Lagrange} A.~M.,  2013, \mn@doi [\aap]
  {10.1051/0004-6361/201219917}, \href
  {https://ui.adsabs.harvard.edu/abs/2013A&A...551A.101M} {551, A101}

\bibitem[\protect\citeauthoryear{{Milbourne} et~al.,}{{Milbourne}
  et~al.}{2021}]{Milbourne2021}
{Milbourne} T.~W.,  et~al., 2021, \apj, 920

\bibitem[\protect\citeauthoryear{{Mortier, }, {Santos, N. C.}, {Sousa, S. G.},
  {Fernandes, J. M.}, {Adibekyan, V. Zh.}, {Delgado Mena, E.}, {Montalto, M.}
  \& {Israelian, G.}}{{Mortier, } et~al.}{2013}]{MortierrefId0}
{Mortier, } {Santos, N. C.} {Sousa, S. G.} {Fernandes, J. M.} {Adibekyan, V.
  Zh.} {Delgado Mena, E.} {Montalto, M.}  {Israelian, G.} 2013, \mn@doi [A\&A]
  {10.1051/0004-6361/201322240}, 558, A106

\bibitem[\protect\citeauthoryear{{Mortier}, {Sousa}, {Adibekyan}, {Brand{\~a}o}
   \& {Santos}}{{Mortier} et~al.}{2014}]{2014A&A...572A..95M}
{Mortier} A.,  {Sousa} S.~G.,  {Adibekyan} V.~Z.,  {Brand{\~a}o} I.~M.,
  {Santos} N.~C.,  2014, \mn@doi [\aap] {10.1051/0004-6361/201424537}, \href
  {https://ui.adsabs.harvard.edu/abs/2014A&A...572A..95M} {572, A95}

\bibitem[\protect\citeauthoryear{{National Academies of Sciences, Engineering,
  and Medicine and others}}{{National Academies of Sciences, Engineering, and
  Medicine and others}}{2018}]{national2018exoplanet}
{National Academies of Sciences, Engineering, and Medicine and others} 2018,
  Exoplanet Science Strategy.
National Academies Press

\bibitem[\protect\citeauthoryear{{Newton} et~al.,}{{Newton}
  et~al.}{2019}]{newton2019}
{Newton} E.~R.,  et~al., 2019, \mn@doi [\apjl] {10.3847/2041-8213/ab2988},
  \href {https://ui.adsabs.harvard.edu/abs/2019ApJ...880L..17N} {880, L17}

\bibitem[\protect\citeauthoryear{{Nicholson} \& {Aigrain}}{{Nicholson} \&
  {Aigrain}}{2022}]{Nicholson2022}
{Nicholson} B.~A.,  {Aigrain} S.,  2022, \mn@doi [\mnras]
  {10.1093/mnras/stac2097}, \href
  {https://ui.adsabs.harvard.edu/abs/2022MNRAS.515.5251N} {515, 5251}

\bibitem[\protect\citeauthoryear{{Owen} \& {Wu}}{{Owen} \&
  {Wu}}{2013}]{2013Owen}
{Owen} J.~E.,  {Wu} Y.,  2013, \mn@doi [\apj] {10.1088/0004-637X/775/2/105},
  \href {https://ui.adsabs.harvard.edu/abs/2013ApJ...775..105O} {775, 105}

\bibitem[\protect\citeauthoryear{{Paxton}, {Bildsten}, {Dotter}, {Herwig},
  {Lesaffre}  \& {Timmes}}{{Paxton} et~al.}{2011}]{2011ApJS..192....3P}
{Paxton} B.,  {Bildsten} L.,  {Dotter} A.,  {Herwig} F.,  {Lesaffre} P.,
  {Timmes} F.,  2011, \mn@doi [\apjs] {10.1088/0067-0049/192/1/3}, \href
  {https://ui.adsabs.harvard.edu/abs/2011ApJS..192....3P} {192, 3}

\bibitem[\protect\citeauthoryear{{Paxton} et~al.,}{{Paxton}
  et~al.}{2013}]{2013ApJS..208....4P}
{Paxton} B.,  et~al., 2013, \mn@doi [\apjs] {10.1088/0067-0049/208/1/4}, \href
  {https://ui.adsabs.harvard.edu/abs/2013ApJS..208....4P} {208, 4}

\bibitem[\protect\citeauthoryear{{Paxton} et~al.,}{{Paxton}
  et~al.}{2015}]{2015ApJS..220...15P}
{Paxton} B.,  et~al., 2015, \mn@doi [\apjs] {10.1088/0067-0049/220/1/15}, \href
  {https://ui.adsabs.harvard.edu/abs/2015ApJS..220...15P} {220, 15}

\bibitem[\protect\citeauthoryear{{Perger} et~al.,}{{Perger}
  et~al.}{2023}]{Perger2023}
{Perger} M.,  et~al., 2023, \mn@doi [\aap] {10.1051/0004-6361/202245092}, \href
  {https://ui.adsabs.harvard.edu/abs/2023A&A...672A.118P} {672, A118}

\bibitem[\protect\citeauthoryear{{Petigura}}{{Petigura}}{2020}]{2020Petigura}
{Petigura} E.~A.,  2020, \mn@doi [\aj] {10.3847/1538-3881/ab9fff}, \href
  {https://ui.adsabs.harvard.edu/abs/2020AJ....160...89P} {160, 89}

\bibitem[\protect\citeauthoryear{{Petigura} et~al.,}{{Petigura}
  et~al.}{2022}]{2022Petigura}
{Petigura} E.~A.,  et~al., 2022, \mn@doi [\aj] {10.3847/1538-3881/ac51e3},
  \href {https://ui.adsabs.harvard.edu/abs/2022AJ....163..179P} {163, 179}

\bibitem[\protect\citeauthoryear{{Queloz} et~al.,}{{Queloz}
  et~al.}{2001}]{2001A&A...379..279Q}
{Queloz} D.,  et~al., 2001, \mn@doi [\aap] {10.1051/0004-6361:20011308}, \href
  {https://ui.adsabs.harvard.edu/abs/2001A&A...379..279Q} {379, 279}

\bibitem[\protect\citeauthoryear{{Queloz} et~al.,}{{Queloz}
  et~al.}{2009}]{Queloz2009}
{Queloz} D.,  et~al., 2009, \mn@doi [\aap] {10.1051/0004-6361/200913096}, \href
  {https://ui.adsabs.harvard.edu/abs/2009A&A...506..303Q} {506, 303}

\bibitem[\protect\citeauthoryear{{Rajpaul}, {Aigrain}, {Osborne}, {Reece}  \&
  {Roberts}}{{Rajpaul} et~al.}{2015}]{2015MNRAS.452.2269R}
{Rajpaul} V.,  {Aigrain} S.,  {Osborne} M.~A.,  {Reece} S.,   {Roberts} S.,
  2015, \mn@doi [\mnras] {10.1093/mnras/stv1428}, \href
  {https://ui.adsabs.harvard.edu/abs/2015MNRAS.452.2269R} {452, 2269}

\bibitem[\protect\citeauthoryear{{Ricker} et~al.,}{{Ricker}
  et~al.}{2015}]{2015JATIS...1a4003R}
{Ricker} G.~R.,  et~al., 2015, \mn@doi [Journal of Astronomical Telescopes,
  Instruments, and Systems] {10.1117/1.JATIS.1.1.014003}, \href
  {https://ui.adsabs.harvard.edu/abs/2015JATIS...1a4003R} {1, 014003}

\bibitem[\protect\citeauthoryear{{Robertson}, {Mahadevan}, {Endl}  \&
  {Roy}}{{Robertson} et~al.}{2014}]{2014Sci...345..440R}
{Robertson} P.,  {Mahadevan} S.,  {Endl} M.,   {Roy} A.,  2014, \mn@doi
  [Science] {10.1126/science.1253253}, \href
  {https://ui.adsabs.harvard.edu/abs/2014Sci...345..440R} {345, 440}

\bibitem[\protect\citeauthoryear{{Rogers}}{{Rogers}}{2015}]{rogers2015}
{Rogers} L.~A.,  2015, \mn@doi [\apj] {10.1088/0004-637X/801/1/41}, \href
  {https://ui.adsabs.harvard.edu/abs/2015ApJ...801...41R} {801, 41}

\bibitem[\protect\citeauthoryear{{Rogers}, {Gupta}, {Owen}  \&
  {Schlichting}}{{Rogers} et~al.}{2021}]{rogers}
{Rogers} J.~G.,  {Gupta} A.,  {Owen} J.~E.,   {Schlichting} H.~E.,  2021,
  \mn@doi [\mnras] {10.1093/mnras/stab2897}, \href
  {https://ui.adsabs.harvard.edu/abs/2021MNRAS.508.5886R} {508, 5886}

\bibitem[\protect\citeauthoryear{{Rossiter}}{{Rossiter}}{1924}]{Rossiter1924}
{Rossiter} R.~A.,  1924, \mn@doi [\apj] {10.1086/142825}, \href
  {https://ui.adsabs.harvard.edu/abs/1924ApJ....60...15R} {60, 15}

\bibitem[\protect\citeauthoryear{{Saar} \& {Fischer}}{{Saar} \&
  {Fischer}}{2000}]{Saar2000}
{Saar} S.~H.,  {Fischer} D.,  2000, \mn@doi [\apjl] {10.1086/312648}, \href
  {https://ui.adsabs.harvard.edu/abs/2000ApJ...534L.105S} {534, L105}

\bibitem[\protect\citeauthoryear{{Saar}, {Butler}  \& {Marcy}}{{Saar}
  et~al.}{1998}]{Saar1998}
{Saar} S.~H.,  {Butler} R.~P.,   {Marcy} G.~W.,  1998, \mn@doi [\apjl]
  {10.1086/311325}, \href
  {https://ui.adsabs.harvard.edu/abs/1998ApJ...498L.153S} {498, L153}

\bibitem[\protect\citeauthoryear{{Scargle}}{{Scargle}}{1982}]{1982ApJ...263..835S}
{Scargle} J.~D.,  1982, \mn@doi [\apj] {10.1086/160554}, \href
  {https://ui.adsabs.harvard.edu/abs/1982ApJ...263..835S} {263, 835}

\bibitem[\protect\citeauthoryear{Seager \& Dotson}{Seager \&
  Dotson}{2010}]{Seager2010}
Seager S.,  Dotson R.,  2010, Exoplanets.
The University of Arizona space science series, University of Arizona Press,
  Tucson

\bibitem[\protect\citeauthoryear{{Seager}, {Kuchner}, {Hier-Majumder}  \&
  {Militzer}}{{Seager} et~al.}{2007}]{2007ApJ...669.1279S}
{Seager} S.,  {Kuchner} M.,  {Hier-Majumder} C.~A.,   {Militzer} B.,  2007,
  \mn@doi [\apj] {10.1086/521346}, \href
  {https://ui.adsabs.harvard.edu/abs/2007ApJ...669.1279S} {669, 1279}

\bibitem[\protect\citeauthoryear{{Shallue} \& {Vanderburg}}{{Shallue} \&
  {Vanderburg}}{2018}]{Shallue2018}
{Shallue} C.~J.,  {Vanderburg} A.,  2018, \mn@doi [\aj]
  {10.3847/1538-3881/aa9e09}, \href
  {https://ui.adsabs.harvard.edu/abs/2018AJ....155...94S} {155, 94}

\bibitem[\protect\citeauthoryear{{Smith} et~al.,}{{Smith}
  et~al.}{2012}]{Smith2012}
{Smith} J.~C.,  et~al., 2012, \mn@doi [\pasp] {10.1086/667697}, \href
  {https://ui.adsabs.harvard.edu/abs/2012PASP..124.1000S} {124, 1000}

\bibitem[\protect\citeauthoryear{{Sneden}}{{Sneden}}{1973}]{1973ApJ...184..839S}
{Sneden} C.,  1973, \mn@doi [\apj] {10.1086/152374}, \href
  {https://ui.adsabs.harvard.edu/abs/1973ApJ...184..839S} {184, 839}

\bibitem[\protect\citeauthoryear{{Sousa}}{{Sousa}}{2014}]{2014dapb.book..297S}
{Sousa} S.~G.,  2014, {ARES + MOOG: A Practical Overview of an Equivalent Width
  (EW) Method to Derive Stellar Parameters}.
pp 297--310, \mn@doi{10.1007/978-3-319-06956-2\_26}

\bibitem[\protect\citeauthoryear{{Sousa}, {Santos}, {Israelian}, {Mayor}  \&
  {Udry}}{{Sousa} et~al.}{2011}]{2011A&A...533A.141S}
{Sousa} S.~G.,  {Santos} N.~C.,  {Israelian} G.,  {Mayor} M.,   {Udry} S.,
  2011, \mn@doi [\aap] {10.1051/0004-6361/201117699}, \href
  {https://ui.adsabs.harvard.edu/abs/2011A&A...533A.141S} {533, A141}

\bibitem[\protect\citeauthoryear{{Sousa}, {Santos}, {Adibekyan}, {Delgado-Mena}
   \& {Israelian}}{{Sousa} et~al.}{2015}]{2015A&A...577A..67S}
{Sousa} S.~G.,  {Santos} N.~C.,  {Adibekyan} V.,  {Delgado-Mena} E.,
  {Israelian} G.,  2015, \mn@doi [\aap] {10.1051/0004-6361/201425463}, \href
  {https://ui.adsabs.harvard.edu/abs/2015A&A...577A..67S} {577, A67}

\bibitem[\protect\citeauthoryear{{Stevenson} et~al.,}{{Stevenson}
  et~al.}{2012}]{Stevenson2012}
{Stevenson} K.~B.,  et~al., 2012, \mn@doi [\apj] {10.1088/0004-637X/754/2/136},
  \href {https://ui.adsabs.harvard.edu/abs/2012ApJ...754..136S} {754, 136}

\bibitem[\protect\citeauthoryear{{Stumpe} et~al.,}{{Stumpe}
  et~al.}{2012}]{Stumpe2012}
{Stumpe} M.~C.,  et~al., 2012, \mn@doi [\pasp] {10.1086/667698}, \href
  {https://ui.adsabs.harvard.edu/abs/2012PASP..124..985S} {124, 985}

\bibitem[\protect\citeauthoryear{{Stumpe}, {Smith}, {Catanzarite}, {Van Cleve},
  {Jenkins}, {Twicken}  \& {Girouard}}{{Stumpe} et~al.}{2014}]{Stumpe2014}
{Stumpe} M.~C.,  {Smith} J.~C.,  {Catanzarite} J.~H.,  {Van Cleve} J.~E.,
  {Jenkins} J.~M.,  {Twicken} J.~D.,   {Girouard} F.~R.,  2014, \mn@doi [\pasp]
  {10.1086/674989}, \href
  {https://ui.adsabs.harvard.edu/abs/2014PASP..126..100S} {126, 100}

\bibitem[\protect\citeauthoryear{{Ter Braak}}{{Ter Braak}}{2006}]{TerBraak2006}
{Ter Braak} C. J.~F.,  2006, \mn@doi [Statistics and Computing]
  {10.1007/s11222-006-8769-1}, \href
  {https://ui.adsabs.harvard.edu/abs/2006S&C....16..239T} {16, 239}

\bibitem[\protect\citeauthoryear{{Thygesen} et~al.,}{{Thygesen}
  et~al.}{2023}]{Thygesen2023}
{Thygesen} E.,  et~al., 2023, \mn@doi [\aj] {10.3847/1538-3881/acaf03}, \href
  {https://ui.adsabs.harvard.edu/abs/2023AJ....165..155T} {165, 155}

\bibitem[\protect\citeauthoryear{{Torres}, {Andersen}  \&
  {Gim{\'e}nez}}{{Torres} et~al.}{2010}]{Torres2010}
{Torres} G.,  {Andersen} J.,   {Gim{\'e}nez} A.,  2010, \mn@doi [\aapr]
  {10.1007/s00159-009-0025-1}, \href
  {https://ui.adsabs.harvard.edu/abs/2010A&ARv..18...67T} {18, 67}

\bibitem[\protect\citeauthoryear{{Tran}, {Bedell}, {Foreman-Mackey}  \&
  {Luger}}{{Tran} et~al.}{2023}]{Tran2023}
{Tran} Q.~H.,  {Bedell} M.,  {Foreman-Mackey} D.,   {Luger} R.,  2023, \mn@doi
  [arXiv e-prints] {10.48550/arXiv.2305.00988}, \href
  {https://ui.adsabs.harvard.edu/abs/2023arXiv230500988T} {p. arXiv:2305.00988}

\bibitem[\protect\citeauthoryear{{Van Cleve} et~al.,}{{Van Cleve}
  et~al.}{2016}]{VanCleve2016}
{Van Cleve} J.~E.,  et~al., 2016, \mn@doi [\pasp]
  {10.1088/1538-3873/128/965/075002}, \href
  {https://ui.adsabs.harvard.edu/abs/2016PASP..128g5002V} {128, 075002}

\bibitem[\protect\citeauthoryear{{Van Eylen}, {Agentoft}, {Lundkvist},
  {Kjeldsen}, {Owen}, {Fulton}, {Petigura}  \& {Snellen}}{{Van Eylen}
  et~al.}{2018}]{2018VanEylen}
{Van Eylen} V.,  {Agentoft} C.,  {Lundkvist} M.~S.,  {Kjeldsen} H.,  {Owen}
  J.~E.,  {Fulton} B.~J.,  {Petigura} E.,   {Snellen} I.,  2018, \mn@doi
  [\mnras] {10.1093/mnras/sty1783}, \href
  {https://ui.adsabs.harvard.edu/abs/2018MNRAS.479.4786V} {479, 4786}

\bibitem[\protect\citeauthoryear{{VanderPlas} \& {Ivezi{\'c}}}{{VanderPlas} \&
  {Ivezi{\'c}}}{2015}]{2015ApJ...812...18V}
{VanderPlas} J.~T.,  {Ivezi{\'c}} {\v{Z}}.,  2015, \mn@doi [\apj]
  {10.1088/0004-637X/812/1/18}, \href
  {https://ui.adsabs.harvard.edu/abs/2015ApJ...812...18V} {812, 18}

\bibitem[\protect\citeauthoryear{{VanderPlas}, {Connolly}, {Ivezic}  \&
  {Gray}}{{VanderPlas} et~al.}{2012}]{2012cidu.conf...47V}
{VanderPlas} J.,  {Connolly} A.~J.,  {Ivezic} Z.,   {Gray} A.,  2012, in
  Proceedings of Conference on Intelligent Data Understanding (CIDU. pp 47--54
  (\mn@eprint {arXiv} {1411.5039}), \mn@doi{10.1109/CIDU.2012.6382200}

\bibitem[\protect\citeauthoryear{Vanderburg}{Vanderburg}{2021}]{andrew_vanderburg_2021_5599854}
Vanderburg A.,  2021, avanderburg/edmcmc: v1.0.0,
  \mn@doi{10.5281/zenodo.5599854}, \url
  {https://doi.org/10.5281/zenodo.5599854}

\bibitem[\protect\citeauthoryear{{Vanderburg} \& {Johnson}}{{Vanderburg} \&
  {Johnson}}{2014a}]{2014PASP..126..948V}
{Vanderburg} A.,  {Johnson} J.~A.,  2014a, \mn@doi [\pasp] {10.1086/678764},
  \href {https://ui.adsabs.harvard.edu/abs/2014PASP..126..948V} {126, 948}

\bibitem[\protect\citeauthoryear{{Vanderburg} \& {Johnson}}{{Vanderburg} \&
  {Johnson}}{2014b}]{Vanderburg2014}
{Vanderburg} A.,  {Johnson} J.~A.,  2014b, \mn@doi [\pasp] {10.1086/678764},
  \href {https://ui.adsabs.harvard.edu/abs/2014PASP..126..948V} {126, 948}

\bibitem[\protect\citeauthoryear{{Vanderburg} et~al.,}{{Vanderburg}
  et~al.}{2016}]{2016ApJS..222...14V}
{Vanderburg} A.,  et~al., 2016, \mn@doi [\apjs] {10.3847/0067-0049/222/1/14},
  \href {https://ui.adsabs.harvard.edu/abs/2016ApJS..222...14V} {222, 14}

\bibitem[\protect\citeauthoryear{{Vanderburg} et~al.,}{{Vanderburg}
  et~al.}{2018}]{vanderburg2018}
{Vanderburg} A.,  et~al., 2018, \mn@doi [\aj] {10.3847/1538-3881/aac894}, \href
  {https://ui.adsabs.harvard.edu/abs/2018AJ....156...46V} {156, 46}

\bibitem[\protect\citeauthoryear{{Weiss} \& {Marcy}}{{Weiss} \&
  {Marcy}}{2014}]{2014ApJ...783L...6W}
{Weiss} L.~M.,  {Marcy} G.~W.,  2014, \mn@doi [\apjl]
  {10.1088/2041-8205/783/1/L6}, \href
  {https://ui.adsabs.harvard.edu/abs/2014ApJ...783L...6W} {783, L6}

\bibitem[\protect\citeauthoryear{{Wise}, {Plavchan}, {Dumusque}, {Cegla}  \&
  {Wright}}{{Wise} et~al.}{2022}]{Wise2022}
{Wise} A.,  {Plavchan} P.,  {Dumusque} X.,  {Cegla} H.,   {Wright} D.,  2022,
  \mn@doi [\apj] {10.3847/1538-4357/ac649b}, \href
  {https://ui.adsabs.harvard.edu/abs/2022ApJ...930..121W} {930, 121}

\bibitem[\protect\citeauthoryear{{Zechmeister} \& {K{\"u}rster}}{{Zechmeister}
  \& {K{\"u}rster}}{2009}]{2009A&A...496..577Z}
{Zechmeister} M.,  {K{\"u}rster} M.,  2009, \mn@doi [\aap]
  {10.1051/0004-6361:200811296}, \href
  {https://ui.adsabs.harvard.edu/abs/2009A&A...496..577Z} {496, 577}

\bibitem[\protect\citeauthoryear{Zeng, Sasselov  \& Jacobsen}{Zeng
  et~al.}{2016}]{Zeng_2016}
Zeng L.,  Sasselov D.~D.,   Jacobsen S.~B.,  2016, \mn@doi [The Astrophysical
  Journal] {10.3847/0004-637x/819/2/127}, 819, 127

\bibitem[\protect\citeauthoryear{{Zhao} et~al.,}{{Zhao}
  et~al.}{2022}]{Zhao2022}
{Zhao} L.~L.,  et~al., 2022, \mn@doi [\aj] {10.3847/1538-3881/ac5176}, \href
  {https://ui.adsabs.harvard.edu/abs/2022AJ....163..171Z} {163, 171}

\bibitem[\protect\citeauthoryear{{de Beurs} et~al.,}{{de Beurs}
  et~al.}{2022}]{deBeurs2022b}
{de Beurs} Z.~L.,  et~al., 2022, \mn@doi [\aj] {10.3847/1538-3881/ac738e},
  \href {https://ui.adsabs.harvard.edu/abs/2022AJ....164...49D} {164, 49}

\makeatother
\end{thebibliography}








\section*{Appendix}
\renewcommand{\thefigure}{A.\arabic{figure}}
\setcounter{figure}{0}

\renewcommand{\thetable}{A.\arabic{table}}
\setcounter{table}{0}

\begin{figure*}
    \includegraphics[width=1.05\linewidth]{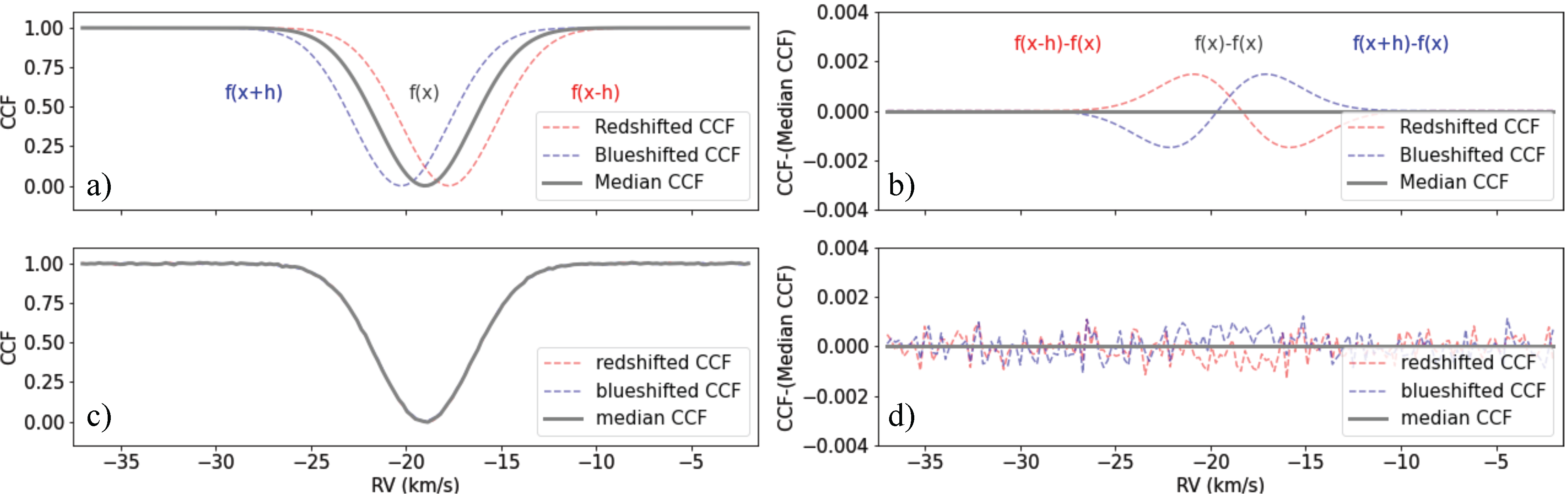}
    \caption{Understanding CCF Shape Diagnostics. a) Simulated CCFs that are red-shifted (red), blue-shifted (blue) and the median CCF (grey) are plotted. These CCFs are not shifted to the center. b) The residual CCF  (simulated CCFs minus the median CCF) are plotted and are \bedit{equivalent} to the derivatives of a Gaussian. c) A red-shifted (red), blue-shifted (blue) and median CCF (grey) are plotted. \bedit{However, in this case } they are all centered at the median RV and we simulated some \bedit{white} Gaussian noise as a \bedit{simple} approximation to shape changes induced by stellar activity. d)  The residual CCF  (simulated CCFs minus the median CCF) are plotted and do not resemble the derivatives of a Gaussian.}
    \label{fig:ccf_shapes}
\end{figure*}

\begin{figure*}
    \includegraphics[width=1.05\textwidth]{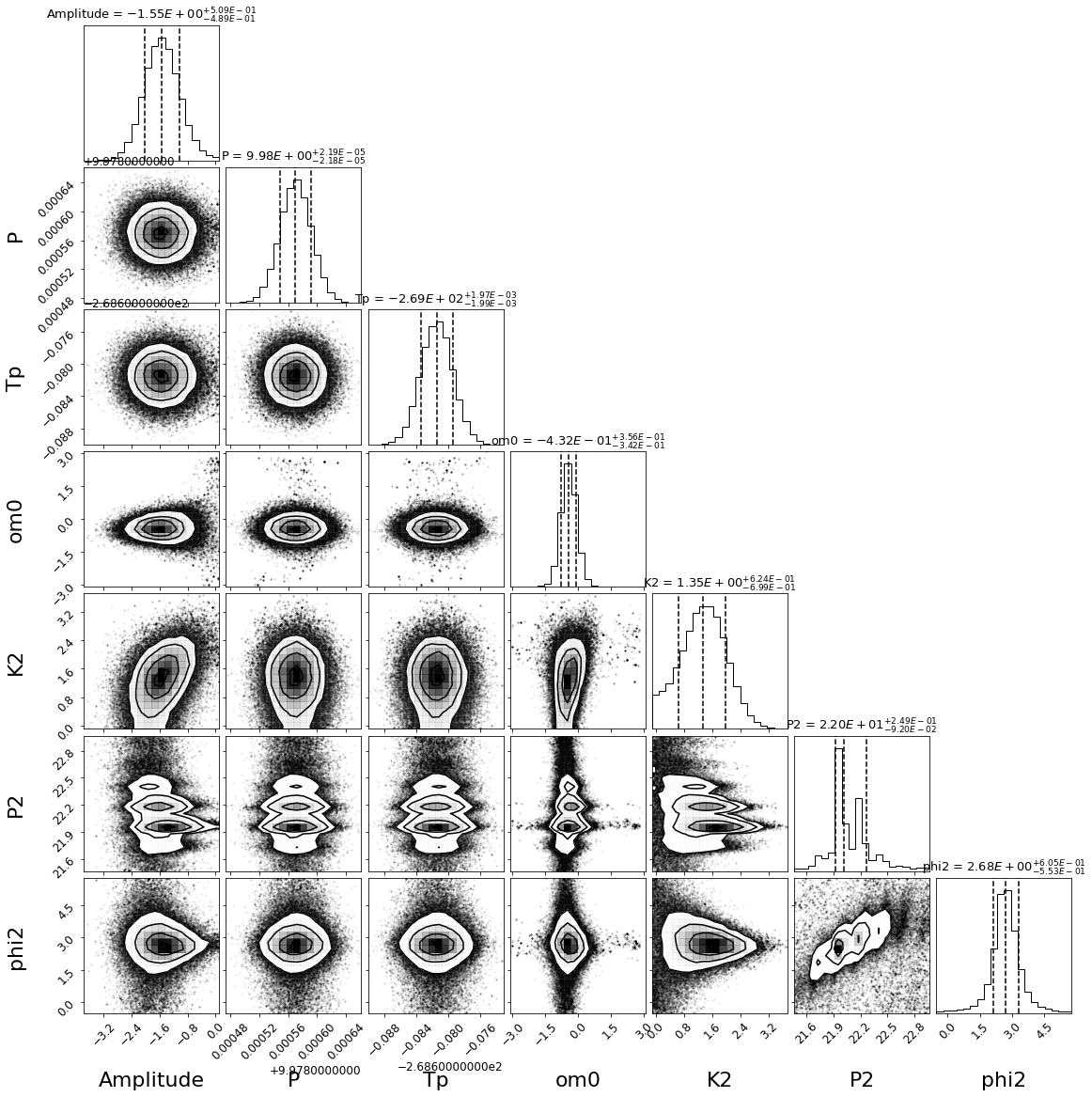}
    \caption{DE-MCMC corner plot for two planet fit. For \thisplanet\ the free parameters are semi-amplitude (Amplitude), period (P), Time of periastron passage (T$_p$). Eccentricity is frozen at e=0 for this fit. The potential companion is modeled by a cosine curve and thus the free parameters are semi-amplitude (K2), period (P2), and phase (phi2). \bedit{For P2, we can clearly see that the data can be consistent with multiple possible periods.}}
    \label{fig:two_planets}
\end{figure*}

\beditrr{In this appendix, we include the radial velocity measurements included in this analysis (Table \ref{harpsrvs}). In addition, we provide a supplemental figure that visually explains the CCF shape diagnostics used to prevent overfitting in the RV analysis (Figure \ref{fig:ccf_shapes}. Lastly, we provide an MCMC corner plot for the two planet RV fit where the the period of a hypothethical companion is poorly constrained (Figure \ref{fig:two_planets}).}

\begin{center}
\begin{singlespace}
\begin{table}
\footnotesize
\begin{tabular}{l|cc}
BJD     & RV [\ms] & RV$_{\rm error}$ [\ms]\\
\hline
\hline
2457618.62  & -17549.45  & 0.89  \\
2457618.55  & -17548.22  & 1.17  \\
2457270.54  & -17555.49  & 0.79  \\
2457269.56  & -17555.69  & 1.2  \\
2457574.74  & -17552.52  & 0.89  \\
. & . & . \\
. & . & . \\
. & . & . \\
2457565.71  & -17553.78  & 1.29  \\
2457562.71  & -17551.56  & 1.2  \\
2457322.45  & -17548.57  & 0.81  \\
2457356.3  & -17548.43  & 1.41  \\
2457354.36  & -17546.95  & 1.26  
\end{tabular}
\caption{\redit{Radial velocity observations and estimated instrumental errors for \thisstar\ using the HARPS-N DRS 2.3.5}. The full table is available online.\label{harpsrvs}}
\end{table}
\end{singlespace}
\end{center}

\bsp	
\label{lastpage}
\end{document}
